\documentclass[useAMS,usenatbib]{mn2e}

\usepackage{graphicx}
\usepackage{txfonts}

\newcommand{\msun}{M$_{\odot}$}
\newcommand{\iso}[2]{\hbox{${}^{#1}{\rm #2}$}}

\newcommand{\daona}{\ensuremath{\Delta\alpha/\alpha \,}}

\newcommand{\kms}{\hbox{${\rm km\,s}^{-1}$}}
\newcommand{\zab}{\hbox{$z_{\rm abs}$}}
\newcommand{\da}{\hbox{$\Delta\alpha/\alpha$}}
\newcommand{\mgirat}{\hbox{$(^{25}{\rm Mg}\!+\!^{26}{\rm Mg})/^{24}{\rm Mg}$}}
\newcommand{\siirat}{\hbox{$(^{29}{\rm Si}\!+\!^{30}{\rm Si})/^{28}{\rm Si}$}}
\newcommand{\bspsmall}{\vspace{0.5cm}\small\noindent This paper has been
typeset from a \TeX / \LaTeX\ file prepared by the author.}

\defcitealias{MurphyM_04a}{MFW04}
\defcitealias{AshenfelterT_04b}{AMO04}

\title[Varying $\bmath{\alpha}$ and AGB pollution]{On variations in the
fine-structure constant and stellar pollution of quasar absorption
systems}
   \author[Y. Fenner, M. T. Murphy, B. K. Gibson]{
     Y. Fenner$^{1}$\thanks{E-mail: yfenner@astro.swin.edu.au, mim@ast.cam.ac.uk },
     M. T. Murphy$^{2}$\footnotemark[1],
     B. K. Gibson$^{1}$\\ 
     $^{1}$Centre for Astrophysics \& Supercomputing, Swinburne University 
       of Technology, Melbourne, Australia\\
     $^{2}$Institute of Astronomy, University of Cambridge, Madingley Road,
       Cambridge CB3 0HA, UK }

\begin{document}

\date{Accepted ---. Received ---; in original form ---}

\pagerange{\pageref{firstpage}--\pageref{lastpage}} \pubyear{2004}

\maketitle

\label{firstpage}

\begin{abstract}    
At redshifts $\zab\!\la\!2$, quasar absorption-line constraints on
space-time variations in the fine-structure constant, $\alpha$, rely on the
comparison of Mg{\sc \,ii} and Fe{\sc \,ii} transition wavelengths. One
potentially important uncertainty is the relative abundance of Mg isotopes
in the absorbers which, if different from solar, can cause spurious shifts
in the measured wavelengths and, therefore, $\alpha$. Here we explore
chemical evolution models with enhanced populations of intermediate-mass
(IM) stars which, in their asymptotic giant branch (AGB) phase, are thought
to be the dominant factories for heavy Mg isotopes at the low metallicities
typical of quasar absorption systems. By design, these models partially
explain recent Keck/HIRES evidence for a smaller $\alpha$ in $\zab\!<\!2$
absorption clouds than on Earth. However, such models also over-produce N,
violating observed abundance trends in high-\zab\ damped Lyman-$\alpha$
systems (DLAs). Our results do not support the recent claim of Ashenfelter,
Mathews \& Olive (2004b) that similar models of IM-enhanced initial mass
functions (IMFs) may simultaneously explain the HIRES varying-$\alpha$ data
and DLA N abundances. We explore the effect of the IM-enhanced model on Si,
Al and P abundances, finding it to be much-less pronounced than for N. We
also show that the $^{13}$C/$^{12}$C ratio, as measured in absorption
systems, could constitute a future diagnostic of non-standard models of the
high-redshift IMF.
\end{abstract} 



\section{Introduction}\label{s:intro}

In the past few years, evidence has emerged that the fine-structure
constant, $\alpha\!\equiv\!e^2/\hbar c$, may have been smaller in
high-redshift quasar (QSO) absorption systems than the value measured today
on Earth \citep[e.g.][hereafter
\citetalias{MurphyM_04a}]{WebbJ_99a,MurphyM_04a}. A possible explanation
for the lower-redshift half of this result is that the abundances of the
heavy Mg isotopes ($^{25}$Mg and $^{26}$Mg) in the absorbers are much
higher, relative to that of $^{24}$Mg, than solar values. Recently,
\citet*{AshenfelterT_04a} proposed and expanded upon \citep*[][hereafter
\citetalias{AshenfelterT_04b}]{AshenfelterT_04b} a chemical evolution model
with an initial mass function (IMF) strongly enhanced at intermediate
masses (IMs), whereby highly super-solar abundances of heavy Mg isotopes
are produced via asymptotic giant branch (AGB) stars. In this paper we
explore the side-effects of this model with a view to identifying possible
observational signatures other than increased heavy Mg isotope abundances.

The paper is organised as follows. The remainder of this section summarizes
the QSO absorption-line evidence for and against variations in $\alpha$ and
describes the sensitivity of that evidence to variations in isotopic
abundances. Section \ref{nucleosynthesis} briefly describes the
nucleosynthesis of Mg isotopes in stars of different masses. In Section
\ref{model} we detail our chemical evolution models and compare them with
those used by \citetalias{AshenfelterT_04b}. Section \ref{results} presents
our main results for the predicted evolution of the total and isotopic
abundances of various elements typically observed in QSO spectra. In each
case we compare these with available data from QSO absorption-line and
local stellar and interstellar medium (ISM) studies and discuss their dominant
uncertainties. Section \ref{discussion} assesses the overall validity of
the IM-enhanced models in light of our new results and presents other
arguments for and against such models. Section \ref{conclusions} gives our
main conclusions.

\subsection{Evidence for varying $\bmath{\alpha}$ from QSO
  absorption systems?}\label{ss:QSO}

The universality and constancy of the laws of nature rely on the space-time
invariance of fundamental constants, such as $\alpha$. Therefore, since
\citet{MilneE_35a,MilneE_37a} and \citet{DiracP_37a} first suggested the
time-variation of the Newton gravitational constant, a great diversity of
theoretical and experimental exploration of possible space-time variations
in fundamental constants has been pursued \citep[e.g.~see review
in][]{UzanJ_03a}.

High resolution spectroscopy of absorption systems lying along the
lines-of-sight to background QSOs has provided particularly interesting
constraints on variations in $\alpha$ over large spatial and temporal
baselines. Early work focused on the alkali-doublet (AD) method: since the
relative wavelength separation between the two transitions of an AD is
proportional to $\alpha^2$ \citep[e.g.][]{BetheH_77a}, comparison between
AD separations seen in absorption systems with those measured in the
laboratory provides a simple probe of $\alpha$ variation. Several authors
(e.g.~\citealt{VarshalovichD_94a,CowieL_95a};
\citealt*{VarshalovichD_96b,VarshalovichD_00a}) applied the AD method to
doublets of several different ionic species (e.g.~C{\sc \,iv}, Si{\sc
\,ii}, Si{\sc \,iv}, Mg{\sc \,ii} and Al{\sc \,iii}). The strongest current
AD constraints come from analysis of many Si{\sc \,iv} absorption systems
in $R\!\sim\!45\,000$ spectra: $\da\!=\!(-0.5 \pm 1.3) \times 10^{-5}$
\citep[][21 systems; $2.0\!<\!\zab\!<\!3.0$]{MurphyM_01c} and
$\da\!=\!(0.15 \pm 0.43) \times 10^{-5}$ \citep[][15 systems;
$1.6\!<\!\zab\!<\!2.9$]{ChandH_04b}\footnote{\da\ is defined as
$\da\!=\!(\alpha_z\!-\!\alpha_0)/\alpha_0$, for $\alpha_z$ and $\alpha_0$
the values of $\alpha$ in the absorption system(s) and in the laboratory
respectively.}

Considerable recent interest has focused on the many-multiplet (MM) method
introduced by \citet*{DzubaV_99b,DzubaV_99a} and \citet{WebbJ_99a}. The MM
method is a generalization of the AD method, constraining changes in
$\alpha$ by utilizing many observed transitions from different multiplets
and different ions associated with each QSO absorption system. It holds
many important advantages over the AD method, including an effective
order-of-magnitude precision gain stemming from the large differences in
sensitivity of light (e.g.~Mg, Si, Al) and heavy (e.g.~Fe, Zn, Cr) ions to
varying $\alpha$. At low redshift ($0.5\!\la\!\zab\!\la\!1.8$), Mg lines,
whose red transition wavelengths ($\lambda\!>\!2700$\,\AA) are relatively
insensitive to changes in $\alpha$, act as anchors against which the larger
expected shifts in the bluer ($2300\!<\!\lambda\!<\!2700$\,\AA) Fe
transition wavelengths can be measured. At higher \zab, transitions from Si
and Al provide anchor lines distributed in wavelength space amongst a
variety of Cr, Fe, Ni and Zn transitions which shift by large amounts in
both positive and negative directions as $\alpha$ varies. This diversity at
high-\zab\ ensures greater reliability in the face of simple systematic
effects compared with the low-\zab\ Mg/Fe systems.

The MM method, applied to Keck/HIRES QSO absorption spectra, has yielded
very surprising results, with the first tentative evidence for a varying
$\alpha$ by \citet{WebbJ_99a} becoming stronger with successively larger
samples (\citealt{MurphyM_01a,WebbJ_01a}; \citealt*{MurphyM_03a}). The most
recent Keck/HIRES constraint comes from 143 absorption systems over the
range $0.2\!<\!\zab\!<\!4.2$ \citepalias{MurphyM_04a}: $\da\!=\!(-0.57 \pm
0.11) \times 10^{-5}$. This result is quite internally robust: it comprises
three different observational samples and approximately equal low- and
high-\zab\ subsamples, all of which give consistent results. Stubbornly, it
has also proven resistant to a range of potential instrumental and
astrophysical systematic effects \citep{MurphyM_01b,MurphyM_03a}.

Intriguingly, \citet{ChandH_04a} \citep[see also][]{SrianandR_04a} have
analysed 23 Mg/Fe absorption systems in higher signal-to-noise ratio (S/N)
spectra {\it from a different telescope and spectrograph}, the VLT/UVES,
claiming a precise, null result over the range $0.4\!<\!\zab\!<\!2.3$:
$\da\!=\!(-0.06 \pm 0.06) \times 10^{-5}$. \citet*{QuastR_04a} and
\citet{LevshakovS_04a} also find null results in individual UVES
absorbers. The discrepancy between the VLT/UVES and Keck/HIRES results is
yet to be resolved. However, it is important to note that low-order
distortions of the Keck/HIRES wavelength scale, as might be expected from
simple instrumental systematic errors, produce opposite effects on the low-
and high-\zab\ samples and so cannot fully explain the HIRES--UVES
difference \citep{MurphyM_03a,MurphyM_04a}. If the HIRES result is
incorrect, the nature of the contributory systematic errors must be subtle
and somewhat conspiratorial.

\subsection{Isotopic abundance evolution?}\label{ss:AGB}

\begin{figure*}
\centering
\hbox{
  \includegraphics[width=6cm,height=8cm]{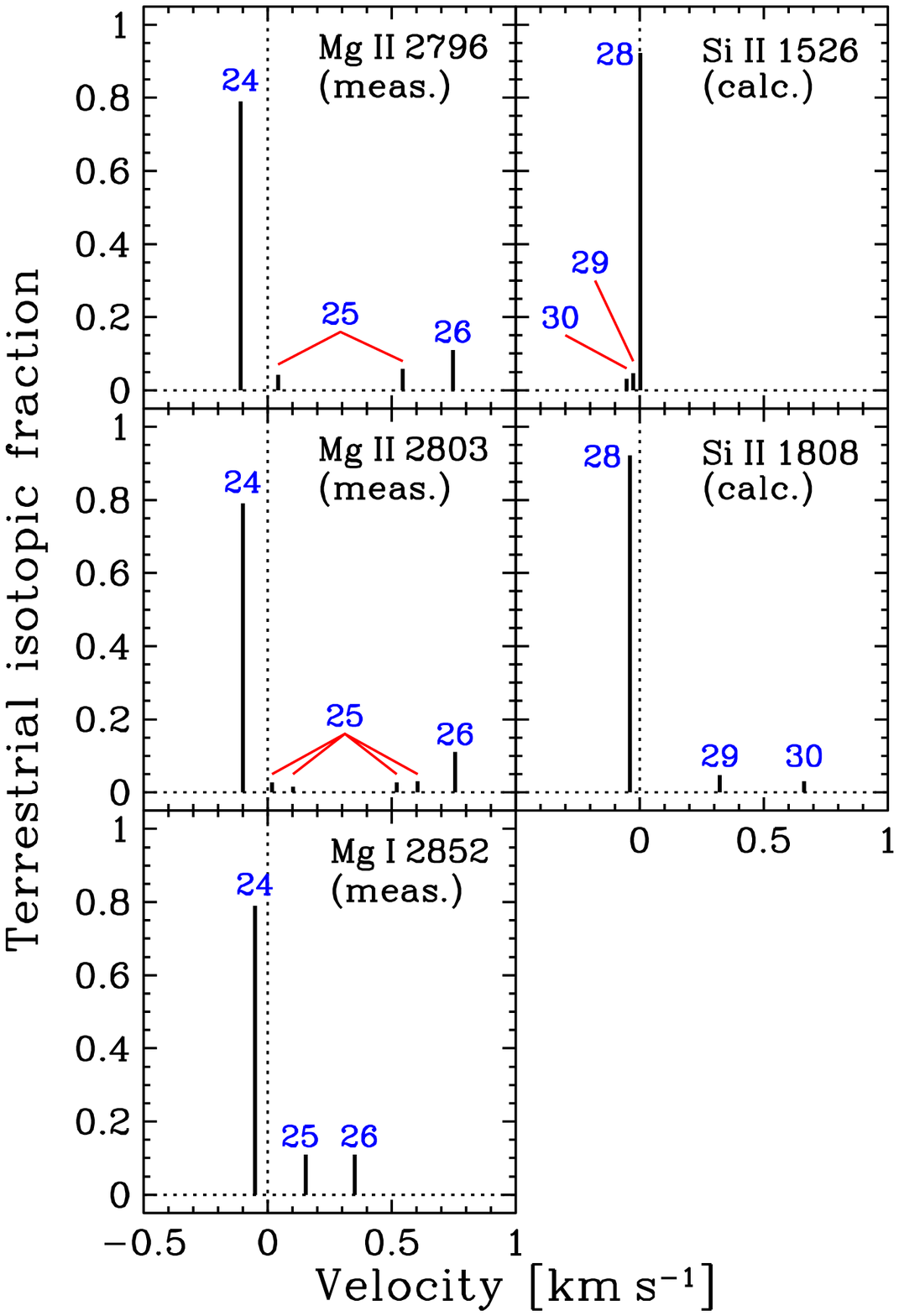}
  \hspace{0.2cm}
  \includegraphics[height=8cm]{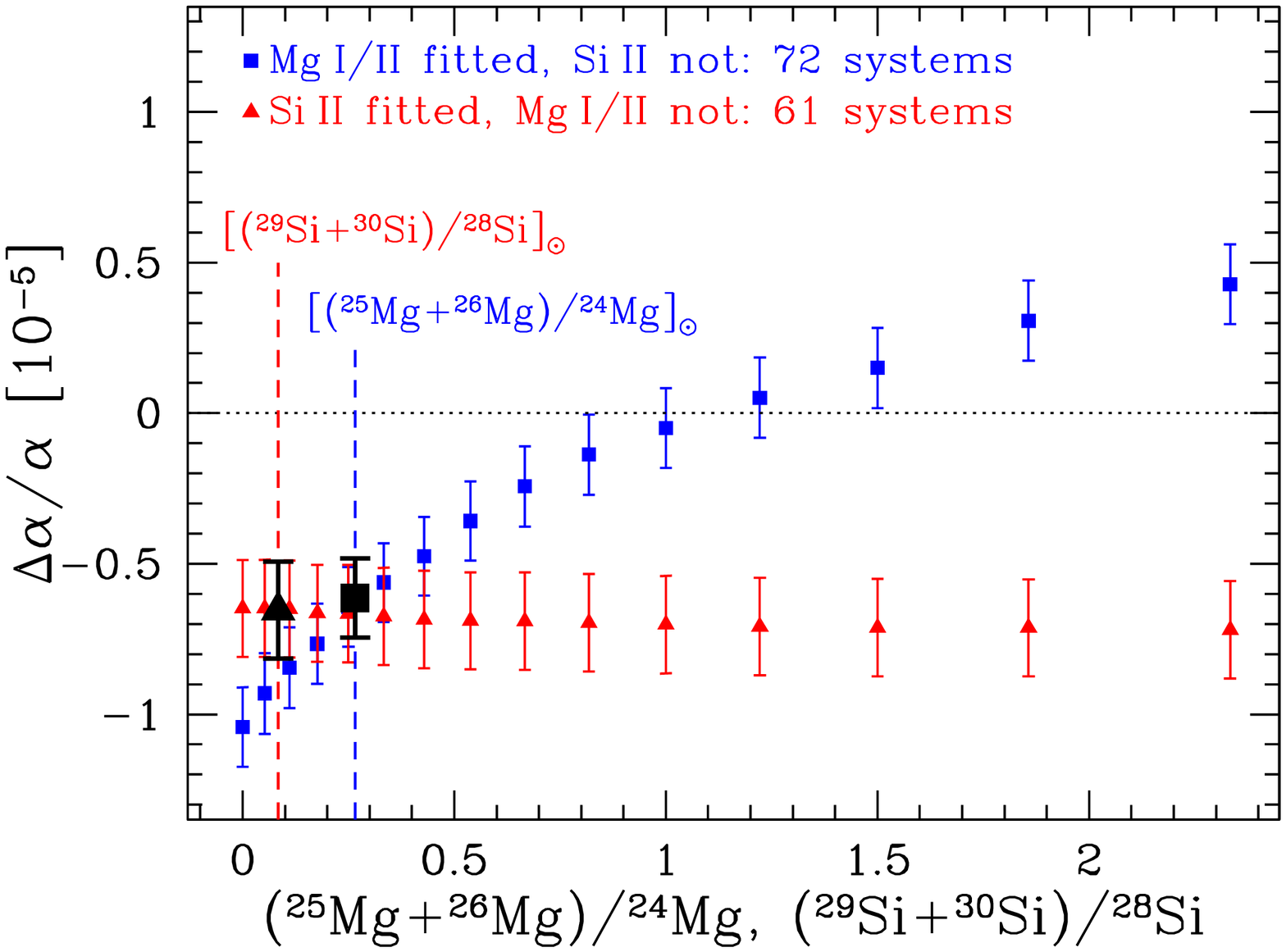}
}
\caption{Left panel: Isotopic structures for the relevant Mg/Si transitions
  from measurements \citep{HallstadiusL_79a,DrullingerR_80a,PickeringJ_98a}
  or calculations \citep{BerengutJ_03a}. Zero velocity corresponds to the
  isotopic structure's centre of gravity. Right panel: Sensitivity of \da\
  measured by \citetalias{MurphyM_04a} to variations in Mg and Si isotopic
  abundance variations. Absorption systems containing no measured Si lines
  (i.e.~most low-\zab\ systems) contribute to the square points while only
  systems containing Si lines and no measured Mg lines (i.e.~most
  high-\zab\ systems) contribute to the triangular points. The large square
  and triangle mark the values of \da\ obtained with terrestrial relative
  isotopic abundances \citep{RosmanK_98a}:
  ($^{24}$Mg\,:\,$^{25}$Mg\,:\,$^{26}$Mg)$_\odot$\,=\,(79\,:\,10\,:\,11)
  and
  ($^{28}$Si\,:\,$^{29}$Si\,:\,$^{30}$Si)$_\odot$\,=\,(92.2\,:\,4.7\,:\,3.1).
  The abundance ratio of the heavy isotopes [i.e.~($^{25}$Mg\,:\,$^{26}$Mg)
  and ($^{29}$Si\,:\,$^{30}$Si)] was held at the terrestrial value along
  the abscissa.}
\label{alpha_iso_fig}
\end{figure*}

\citet{MurphyM_01b} first identified the potential systematic error
introduced into the MM method if the relative isotopic abundances of
crucial anchor elements like Mg and Si underwent strong cosmological
evolution. This is because the absorption lines of the different isotopes
are spaced widely enough ($\sim\!0.5\,\kms$) in these light ions to affect
the measured line centroids (Fig.~\ref{alpha_iso_fig}, left
panel). \citetalias{MurphyM_04a} calculated \da\ from their HIRES
absorption line data as a function of the assumed relative isotopic
abundances of Mg and Si. The results are presented in the right panel of
Fig.~\ref{alpha_iso_fig}. Note the relative insensitivity of \da\ in the
high-\zab\ systems to the Si heavy isotope ratio, \siirat. As noted above,
this is expected because of the greater diversity of transitions and
line-shifts available in the high-\zab\ regime. It is therefore important
to note that evolution in the Si isotopic abundance cannot explain the
high-\zab\ HIRES results. Until recently, no information about the isotopic
shifts in transitions from heavier ions used at high-\zab\ (e.g.~Cr{\sc
\,ii}, Fe{\sc \,ii}, Ni{\sc \,ii}, Zn{\sc \,ii}) were available. However,
due to their higher masses, the shifts are expected to be smaller with
respect to the sensitivity of the lines to varying $\alpha$. Indeed, the
recent isotopic shift calculations of \citet{KozlovM_04a} confirm this.

\begin{figure}
\centering
\includegraphics[width=8.6cm]{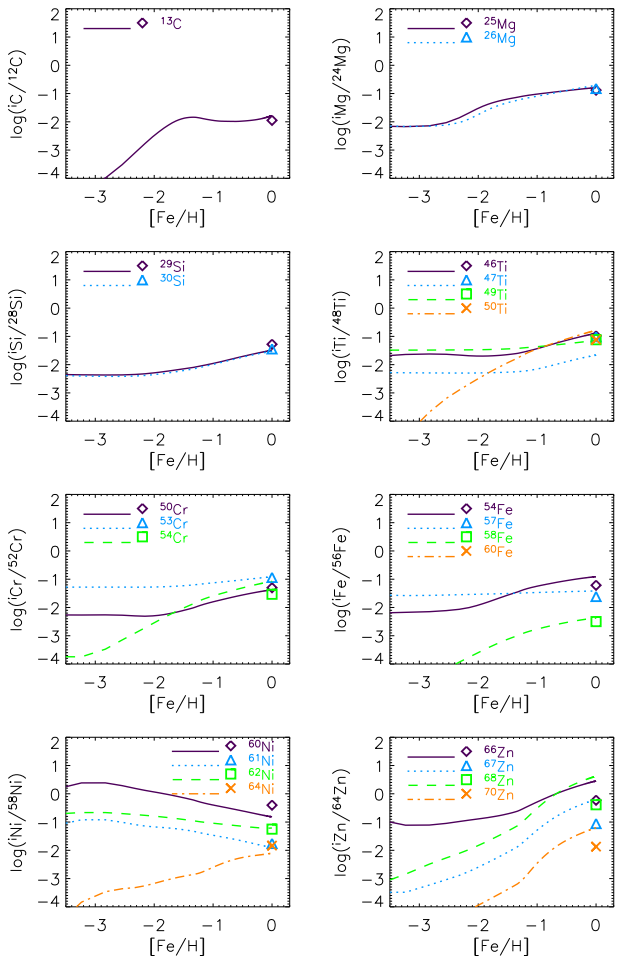}
      \caption{The evolution of isotopic ratios with metallicity
      [Fe/H] for the elements C, Mg, Si, Ti, Cr, Fe, Ni, Zn, which are
      used in varying {\daona} analyses. Isotopic abundances are shown
      on a logarithmic scale relative to the dominant isotope for that
      element. Different curves illustrate the predicted behaviour of
      the various isotopes from a standard solar radius chemical
      evolution model. Symbols show the corresponding solar values. In
      general, the minor isotopes accrue more slowly in the ISM with
      respect to the dominant isotope for each element. A notable
      exception is Ni, for which $^{60,61,62}$Ni/$^{58}$Ni declines
      over time. The departure of Ni from the general rule of thumb
      may reflect uncertainties in its nucleosynthesis, including an
      overproduction from SNe~Ia, that have been pointed out in the
      literature (e.g. Timmes et al. 1995; Iwamoto et al. 1999). } 
      \label{iso_grid_fig}
\end{figure}

However, Fig.~\ref{alpha_iso_fig} emphasises the strong dependence of \da\
in the low-\zab\ systems on the Mg heavy isotope ratio, \mgirat. The
importance of this potential systematic effect therefore depends entirely
on the evolution of Mg isotopic ratios in QSO absorption
systems. Unfortunately, no direct measurement of \mgirat\ in QSO absorbers
is currently feasible due to the small separation of the isotopic
absorption lines. Fortunately however, Mg is one of the few elements for
which stellar isotopic abundances can be measured through molecular
absorption lines, in this case transitions of MgH. \citet{GayP_00a} and
\citet*{YongD_03b} have shown that observed stellar values of \mgirat\
generally decrease with decreasing [Fe/H], as predicted in the
Galactic chemical evolution models of \citet*{TimmesF_95a}. The
(normal IMF) model we present in Section \ref{model} also predicts
such a decrease (Fig.~\ref{iso_grid_fig}). This has been used to argue
that in the low-metallicity environments of Mg/Fe absorbers
(typically, ${\rm [Zn/H]}\!\sim\!-1.0$), one should expect sub-solar
values of \mgirat\ and so, if anything, one expects the low-\zab\
values of \da\ to be {\it too positive}
\citep{MurphyM_01b,MurphyM_03a,MurphyM_04a,ChandH_04a}.

Contrary to this trend, \citet{YongD_03a} have found very high values of
\mgirat\ for some giant stars in the globular cluster NGC 6752, which has
metallicity ${\rm [Fe/H]}\!\sim\!-1.6$. One star had
$\mgirat\!=\!0.91$ (cf.~$[\mgirat]_\odot\!=\!0.27$). Since IM AGB
stars are thought to produce significant quantities of $^{25}$Mg and
$^{26}$Mg (see Section~\ref{nucleosynthesis}),
\citet{YongD_03a} proposed that low-metallicity, IM AGB stars may have
polluted this globular cluster. This prompted
\citet{AshenfelterT_04a,AshenfelterT_04b} to propose a chemical evolution
model with a strongly enhanced population of IM stars as a possible
explanation for the low-\zab\ HIRES varying-$\alpha$ results. In the following
sections we describe in detail the nucleosynthesis of Mg isotopes and
construct a chemical evolution model, similar to that of
\citetalias{AshenfelterT_04b}, to investigate the various effects of an
IM-enhanced IMF.

\section{Nucleosynthesis of magnesium isotopes}\label{nucleosynthesis}

Massive stars culminating in Type II SNe are responsible for most of the Mg
isotopes in the present-day Galaxy. However, there is evidence that
intermediate-mass stars may dominate the production of the neutron-rich Mg
isotopes in the metal-poor regime. According to standard models of stellar
nucleosynthesis, the yield of the heavy Mg isotopes, \iso{25,26}{Mg},
scales with the initial stellar metallicity.  Conversely, the generation of
\iso{24}{Mg} from SNe~II operates fairly independently of initial
metallicity. Since massive stars alone are insufficient to account for the
higher than expected values of \iso{25,26}{Mg}/\iso{24}{Mg} detected in
metal-poor stars (Gay \& Lambert 2000; Yong et al. 2003b), it has been
suggested that there is a supplemental source of the neutron-rich Mg
isotopes.

Karakas \& Lattanzio (2003) have shown that \iso{25}{Mg} and \iso{26}{Mg}
production is substantial in metal-poor intermediate-mass stars (IMSs). At
low metallicities, asymptotic giant branch stars are believed to generate
\iso{25}{Mg} and \iso{26}{Mg} from $\alpha$-capture onto $^{22}$Ne
triggered by He-shell thermal pulsing.  Temperatures at the base of the
convective envelope in 4--6\,\msun\ stars can be high enough to burn
\iso{24}{Mg} via hot bottom burning (HBB) as well as synthesise large
amounts of \iso{25}{Mg} and \iso{26}{Mg}. Based on the recent
nucleosynthetic calculations from Karakas \& Lattanzio (2003), AGB stars
have been shown to produce sufficient quantities of \iso{25,26}{Mg} to
resolve the discrepancy between local stellar observations and previous
model predictions (Fenner et al. 2003). For the present study, we have
incorporated the same grids of low- and intermediate-mass stellar yields
that successfully reproduced the solar neighbourhood Mg isotopic evolution.


\section{Chemical evolution models}\label{model}

The distribution of elements and isotopes as a function of time and radius
was simulated for a Milky Way-like disk galaxy by numerically solving the
classic set of equations (as described by Tinsley 1980) governing gas
infall, star formation, stellar evolution and nucleosynthesis. In
particular, we present predictions for (i) the solar annulus ($r =$ 8.5
kpc) because it is the solar neighbourhood for which we have the most
comprehensive set of empirical constraints, and (ii) the outer disk ($r =$
16.5 kpc) because it is possible that many QSO absorption systems
correspond to the outer regions of spiral disks (e.g. Dessauges-Zavadsky et
al. 2004).  To summarise the details of the model: we define
$\sigma_i(r,t)$ as the mass surface density of species $i$ at time $t$ and
radius $r$, and assume that its rate of change of is given by:

\begin{eqnarray}
\displaystyle\frac{d}{dt} \sigma_{i} (r,t) & = & \displaystyle\int^{m_{\rm
 up}}_{m_{\rm low}}\psi(r,t-\tau_{m})\,Y_i(m,Z(r,t-\tau_{m}))
 \,\frac{\displaystyle\phi(m)}{m} \,\; dm \nonumber \\ & & +
 \displaystyle\frac{d}{dt} \sigma_i(r,t)_{\rm infall} -
 X_i(r,t)\,\psi(r,t),
\end{eqnarray}

\noindent
where the three terms on the right-hand side correspond to the stellar
ejecta, gas infall, and star formation, respectively. The star
formation rate, $\psi$, varies with the square of the gas surface
density in our models, consistent with the empirical Schmidt (1959)
law. $Y_i(m,Z(r,t-\tau_{m}))$ denotes the stellar yield of $i$ (in
mass units) from a star of mass $m$ and metallicity $Z(r,t-\tau_{m})$,
$\phi(m)$ is the initial mass function, and $X_i$ is the mass fraction
of element $i$. By definition, the sum of $X_i$ over all $i$ is
unity. The total surface mass density is identical to the integral
over the infall rate. The lower and upper stellar mass limits, $m_{\rm
low}$ and $m_{\rm up}$, are 0.08 {\msun} and 60 {\msun}, respectively,
while $\tau_{m}$ is the main-sequence lifetime of a star of mass
$m$. We split the first term into three equations that deal separately
with low- and intermediate-mass stars, Type~Ia supernova (SN)
progenitors, and massive stars.

Further details of the numerical code employed in this study can be
found in Fenner \& Gibson (2003), Fenner et al. (2003) and Fenner,
Prochaska \& Gibson (2004).

\subsection{Infall scheme}

We assumed that the Milky Way-like disk galaxy formed during two main
gas accretion episodes. The first occurs on a rapid timescale ($<$ 0.5
Gyr) and is associated with the formation of the halo and thick disk,
while the second episode occurs on a longer timescale and fuels the
formation of stars in the disk. For simplicity, we have assumed no
prior metal-enrichment of the gas infalling onto the disk, although
there is some evidence from observations of high-velocity clouds that
gas falling into the Galaxy may contain traces of heavy elements
(e.g. Wakker et~al. 1999; Gibson et~al. 2001; Sembach et~al. 2002).
Exponentially decaying infall rates have been adopted, such that the
evolution of total surface mass $\sigma_{tot}(r,t)$ density is given
by
\begin{equation}
\frac{d \sigma_{\rm tot}(r,t)}{dt} = A(r) e^{-t/\tau _{\rm H}(r)} + B(r)
e^{-(t-t_{\rm delay})/ \tau _{\rm D}(r)}
\end{equation}
\noindent
where the infall rate coefficients $A(r)$ and $B(r)$ are chosen in
order to reproduce the present-day surface mass density of the
halo/thick disk and thin disk components, which we take to be 10 and
45\,M$_{\odot}$\,pc$^{-2}$, respectively. The adopted timescales for
the infall phases are $\tau_{\rm H}$\,=\,0.1\,Gyr and $\tau_{\rm
D}$\,=\,9.0\,Gyr at the solar radius $r_{\odot}$\,=\,8.5\,kpc. Disk
formation starts after an initial delay $t_{\rm delay} = 0.5$\,Gyr.
The `inside-out' functional form for $\tau_{\rm D}(r)$ (Romano
et~al. 2000) is adopted, whereby the timescale for disk gas accretion
increases linearly with radius. The Milky Way age is taken to be
13\,Gyr.

\subsection{Initial mass function (IMF)}\label{imf}

\begin{figure}
\centering
\includegraphics[width=8cm]{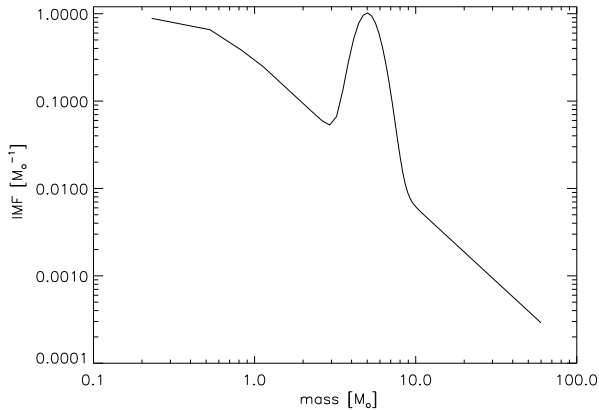}
      \caption{The shape of the IM-enhanced IMF used in this study at
      time $t$ = 0. The IMF is composed of an AGB-biased bump
      superimposed on a standard Kroupa, Tout \& Gilmore (1993)
      3-component power-law. The height and width of the AGB bump was
      chosen to emulate that in \citetalias{AshenfelterT_04b}. The
      amplitude of the bump decays exponentially with increasing gas
      metallicity.}
      \label{imf_fig}
\end{figure}

The shape of the stellar initial mass function (IMF) influences the
quantity of Galactic material locked up in stars of different masses, which
in turn determines the rate at which different elements are released into
the ISM.  The models presented in Section~\ref{results} compare the Kroupa,
Tout \& Gilmore (1993) three-component IMF with one enhanced in AGB stars
at low metallicities. The shape of the IM-enhanced IMF at time $t = 0$ is
illustrated in Fig.~\ref{imf_fig}. The height and width of the AGB bump
was chosen to approximate that in \citetalias{AshenfelterT_04b}. It is
centred on 5 {\msun} with a narrow mass range, although not as narrow as
the \citetalias{AshenfelterT_04b} Model 1, for which the extra AGB
population consists almost entirely of 5$\pm$1~\msun\ stars (compare our
Fig.~\ref{imf_fig} with their figure~3). Their preferred models
(i.e. Models 2 and 3) adopt a wider IM-bump and we also favour a broader
peak, on the grounds that it is more physically realistic\footnote{As a
check, we ran a model using a narrower and taller IM-bump, conserving the
mass contained in the AGB-enhancement.  The final results were similar in
both cases.}.

In our IM-enhanced models, the amplitude of the IM-bump decays
exponentially with increasing gas metallicity.  This differs from
\citetalias{AshenfelterT_04b}, who adopted time-dependent decay.  Imposing
a metallicity- rather than time-dependent IM-enhancement was designed to
reflect the different physical properties of metal-poor gas clouds. The
decreased cooling rate and magnetic field strength in metal-poor material
is expected to influence the mass distribution of newly formed stars. For
consistency with \citetalias{AshenfelterT_04b} Model 1, our scale-factor
for the exponentially decaying IM-bump was chosen to correspond to the
predicted metallicity at the solar radius at time = 0.2 Gyr. This
corresponds to ${\rm [Z/H]}\!\sim\!-1.3$, or roughly 1/20th solar
metallicity. The transformation from a time- to a metallicity-dependent IMS
burst makes it reasonable to apply the same IMF formalism across all
Galactic radii. Because the outer disk has a more protracted star formation
rate and builds up metals more slowly than in the solar radius, the effects
of an IMS burst would be suppressed by adopting a fixed timescale. Instead,
conversion to metallicity-dependence prolongs the temporal duration of the
IMS burst at outer radii and amplifies the chemical signature of an
IM-enhanced IMF.

\subsection{Stellar Yields and Lifetimes}\label{sect_yields}

\emph{Low and intermediate mass stars (LIMS):} For stars less massive than
8 {\msun}, we incorporated the yields from the stellar evolution and
nucleosynthesis code described in Karakas \& Lattanzio (2003), supplemented
with yields for additional isotopes from hydrogen to sulfur and for the
iron-peak, as well as unpublished yields for the metallicity
$Z$=0.0001. The wide range of heavy elements and isotopes incorporated in
these AGB nucleosynthesis models allows us to self-consistently predict the
contribution from low- and intermediate-mass stars (LIMS) to the elemental
and isotopic abundance of many species of interest in DLA and varying
$\alpha$ studies. These include C, N, Mg, Al, Si, Ni and Fe. We also ran
identical chemical evolution models using two alternative sets of LIMS
yields: those of (i) van den Hoek \& Groenewegen (1997) and (ii) Marigo
(2001) (using mixing length parameter $\alpha = 2$ for consistency with
\citetalias{AshenfelterT_04b}). These two alternative inputs do not predict
yields for elements beyond O, but they provide an indication of the
uncertainty in the CNO yields, as will be discussed in
Section~\ref{uncertainties}.

\emph{Type~Ia supernovae (SNe~Ia):} We adopted a recalculation of the
Thielemann, Nomoto, \& Yokoi (1986) W7 model by Iwamoto et~al. (1999) to
estimate the yields from SNe~Ia.  It was assumed that 4\% of binary systems
involving intermediate and low mass stars result in SNe~Ia, since this
fraction provides a good fit to the solar neighbourhood (e.g.  Alib{\' e}s,
Labay, \& Canal 2001; Fenner \& Gibson 2003).

\emph{Massive stars:} For stars more massive than 8--10~{\msun} that end
their lives in violent supernova explosions, we implemented the yields from
Woosley \& Weaver (1995). Their yields span metallicities from zero to
solar and cover the mass range 11--40\,M$_{\odot}$. Since the upper mass
limit of our stellar IMF extends beyond this mass range, we extrapolated
the 35 and 40 {\msun} yields for $m >$ 40 {\msun}. We took the lower energy
`A' models for stars $\le 25$\,M$_{\odot}$ and the higher energy `B' models
for heavier stars. Taking note of the suggestion by Timmes, Woosley, \&
Weaver (1995) that the WW95 mass-cuts may have penetrated too deeply within
the iron core, we have uniformly halved the iron yields from these
models. In the uncertain mass range 8--12 {\msun}, we make the
conservative assumption that these stars do not synthesise new heavy
elements, but expel material with the same heavy element abundance pattern
as the gas from which they were born.

\emph{Stellar lifetimes:} We adopt metallicity-dependent main-sequence
lifetimes calculated by Schaller et~al. (1992). Although stars lose
material over the course of their evolution via stellar winds and planetary
nebulae, this model assumes that all the mass loss takes place at the end
of the main-sequence phase. Our predictions are not significantly affected
by this simplification.

\subsection{Comparison with Ashenfelter et al. (2004)}

We now list some key differences between our model and that of
Ashenfelter et al. (2004b) and describe the impact on the results:

\begin{itemize}

 \item \textbf{AGB yields:} We have adopted the low- and
 intermediate-mass stellar yields from Karakas \& Lattanzio (2003),
 with additional yields for isotopes  up to S as well as the
 Fe-peak, along with unpublished yields for the metallicity
 $Z$=0.0001. Thus, the AGB yields of species such as N and the Mg
 isotopes are internally self-consistent.
 \citetalias{AshenfelterT_04b} added the Mg isotopic yields from
 Karakas \& Lattanzio (2003) onto the Marigo (2001) C, N and O
 yields. Although this means that their N yields were not drawn from
 the same AGB models as the Mg yields, the predicted evolution of N is
 very similar in either case, as discussed in
 Section~\ref{cn_uncertainties}.

 \item \textbf{Type Ia SNe yields:} \citetalias{AshenfelterT_04b}
 employed a metallicity-dependent Type Ia supernovae rate from
 Kobayashi, Tsujimoto \& Nomoto (2000) that prohibits the formation of
 SNe~Ia below a minimum [Fe/H] threshold of $-$1.1. This
 metallicity-dependence was postulated mostly on theoretical
 grounds. Observationally however, $\alpha$/Fe ratios in dwarf
 galaxies (e.g. Shetrone et al. 2001) and S/Zn ratios in DLAs
 (Pettini, Ellison, Steidel \& Bowen 1999) provide evidence for SN~Ia
 activity below [Fe/H] $\sim$ $-$1. Furthermore, the trend of [O/Fe]
 with [Fe/H] seen in local stars can be satisfactorily explained
 without imposing a SN~Ia metallicity threshold (e.g. Alib{\' e}s et
 al. 2001). Thus, the present study does not impose a metallicity
 threshold and the SN~Ia rate is calculated following the method of
 Greggio \& Renzini (1983) and Matteucci \& Greggio (1986). While a
 [Fe/H]~$\sim$~$-$1 SN~Ia threshold would have only a minor effect on
 the solar radius results, the outer disk remains metal-poor for at
 least several Gyr and would be more sensitive to the precise SN~Ia
 prescription, as will be discussed in Section~\ref{results}.

 \item \textbf{Standard IMF:} In recalculating the model of Timmes et
 al. (1995), \citetalias{AshenfelterT_04b} adopted the Salpeter (1955)
 single power-law IMF. We adopt a three-component Kroupa et al. (1993) law,
 which is flatter at lower masses and steeper at the high end of the
 IMF. This empirically derived function leads to a lower overall effective
 yield that is more consistent with local data (Pagel 2001). Although the
 choice of IMF does influence the evolution of the chemical species of
 interest in this investigation, the impact is not large enough to affect
 our final conclusions. Model 1 from \citetalias{AshenfelterT_04b} also
 differs from our IM-enhanced model through the inclusion of an exponential
 0.5 Gyr time scale before the onset of the ``normal'' stellar IMF
 component. Thus, at the very earliest times, most of the star-forming gas
 ends up inside stars with $m \sim$ 5 {\msun}. In contrast, we retain our
 normal IMF component at all times (akin to \citetalias{AshenfelterT_04b}'s
 Model 2).

 \item \textbf{IM-enhanced IMF:} While the
 \citetalias{AshenfelterT_04b} IM-bump decays exponentially on a
 fixed timescale (ranging from 0.2 to 0.4 Gyr depending on the model),
 we transformed this from a time to a metallicity dependence. As
 discussed in Section~\ref{imf}, the assumption that the shape of the
 IMF is governed primarily by the chemical composition of the star
 forming gas cloud is presumably more physically justified than
 imposing a uniform time-dependence.  Converting from the time to the
 metallicity domain allows us to apply the same IMF prescription to
 systems with various star formation histories, such as the slowly
 evolving outer Galactic disk, which we investigate in this paper.

 \item \textbf{Dual-phase infall:} \citetalias{AshenfelterT_04b}
 assumed that the Galactic disk forms during a \emph{single} phase of
 gas infall, in order to directly compare with the results of Timmes
 et al. (1995). We model the Milky Way formation using a
 \emph{dual}-phase infall scheme, since this has been shown to provide
 a better match than single-infall models to the number distribution
 of G- and K-dwarfs in the solar neighbourhood (e.g. Chiappini et
 al. 1997). We stress that our final conclusions are largely
 insensitive to our choice of infall scheme.

\end{itemize}

\section{Results}\label{results}

We now present the predicted chemical evolution for the solar radius
and the outer Galactic disk in the case of an enhanced population of
intermediate-mass stars at low metallicity. These results are compared
with those obtained using a normal IMF and plotted against
observations of local stars and DLAs, where applicable. In all the
following figures, solid and dashed lines denote the evolution at the
solar (8.5 kpc) and outer radii (16.5 kpc), respectively. Models
adopting a normal IMF are shown with thin lines, while thick lines
correspond to models with the IM-enhanced IMF, as illustrated in
Fig.~\ref{imf_fig}.

\subsection{Type Ia supernova rate and the age-metallicity relationship}\label{anr_amr_results}

\begin{figure}
\centering
\includegraphics[width=8cm]{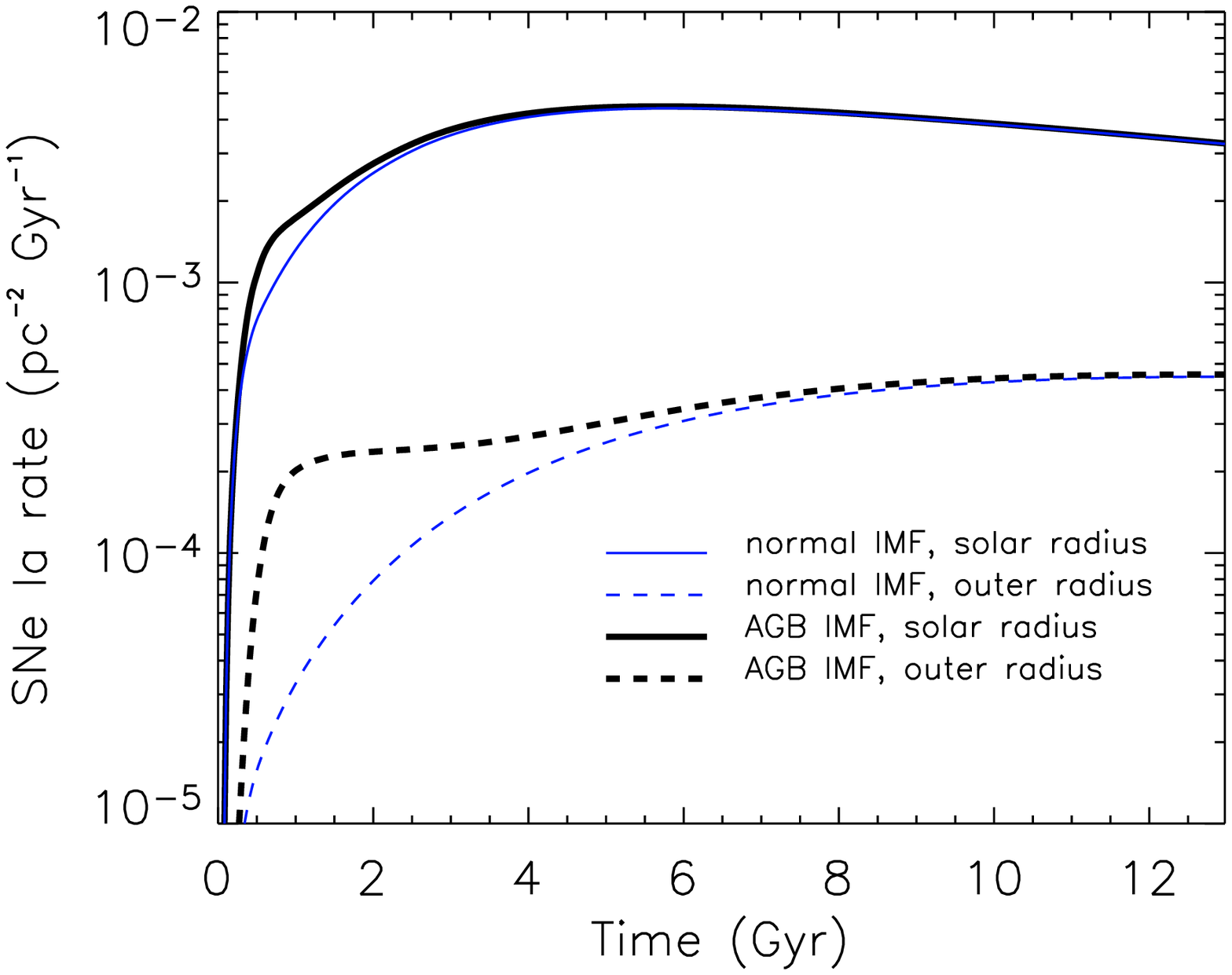}
\includegraphics[width=8cm]{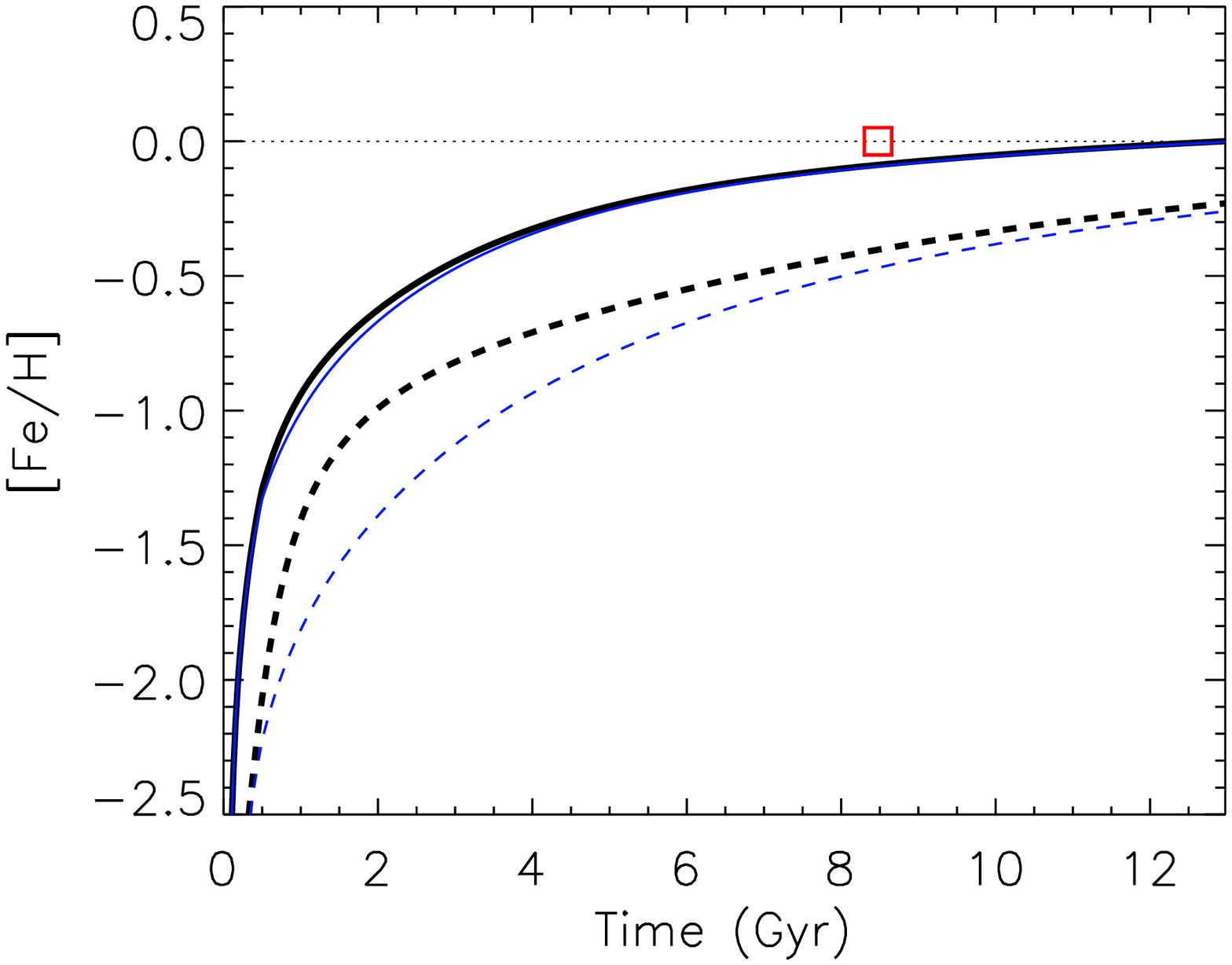}
      \caption{\emph{Upper panel:} Type Ia supernovae rate as a
      function of time for the solar neighbourhood (solid lines) and
      the outer disk (dashed lines) in the case of a normal IMF (thin
      lines) versus the IM-enhanced IMF (thick lines). \emph{Lower
      panel:} Trend of [Fe/H] vs time. The square shows the position
      of the sun, while lines have the same meaning as in the panel
      above. Due to a faster build-up of metals in the solar radius
      relative to the outer disk, the IM-bump decays more quickly and
      leads to only a minor increase in the SN~Ia rate and,
      consequently, the [Fe/H] evolution. In contrast, the slower rate
      of enrichment in the outer radius leads to a longer-lasting
      IM-bump, which has a significant impact on the SN~Ia rate and Fe
      content at early times.}
      \label{snia_fig}
\end{figure}

The evolution of the Type Ia supernova rate predicted by the
model described in Section~\ref{model} is shown in the upper panel of
Fig.~\ref{snia_fig}. The incidence of Type Ia SNe in the solar
region (solid lines) is predicted to have risen to a peak about 7 Gyr
ago and steadily declined thereafter. In contrast, the predicted SN~Ia
rate in the outer disk (dashed lines) has continued to increase up to
the present-day. This difference reflects the more rapid and efficient
conversion of gas into stars in regions of higher surface density.

A consequence of an IMF that increases the number of intermediate-mass
stars is the birth of more Type Ia SNe progenitors. Type Ia SNe are
understood to be associated with binary systems of low- and
intermediate-mass stars, in which the mass lost by the more evolved star is
accreted by its white dwarf (WD) companion until the WD can no longer be
sustained by electron degenerate pressure and a violent explosion
ensues. Comparing the thick and thin solid lines in the upper panel of
Fig.~\ref{snia_fig}, it can be seen that the IM-enhanced IMF has no
significant affect on the local solar neighbourhood SN~Ia rate. In the
outer disk, however, the AGB-enhancement elevates the SN~Ia rate by up to a
factor of six between 0.5--1 Gyr and about a factor two at 3 Gyr before
converging with the normal IMF model after $\sim$6 Gyr (compare the thick
and thin dashed lines).  The different impact of the IM-enhanced IMF on the
solar and outer radii is due to a faster build-up of metals in the solar
radius relative to the outer disk. This causes the IM-bump to decay more
quickly and leads to only a minor increase in the SN~Ia rate. In contrast,
the slower rate of enrichment in the outer radius leads to a longer-lasting
IM-bump, which has a significant impact on the SN~Ia rate at early
times. IM-enhanced IMFs at low metallicity could leave an observable trace
on the cosmic SN~Ia rate, which could be used to discriminate between these
models (Fields et al. 2001).

Type Ia SNe are major producers of iron and are responsible for 1/3--2/3 of
the solar Fe content (e.g. Timmes et al. 1995).  Thus, any increased
incidence of SN~Ia leads inevitably to the production of more Fe, as seen
in the lower panel of Fig.~\ref{snia_fig} where the evolution of [Fe/H]
is plotted. The inclusion of an enhanced IM AGB population increases the
abundance of iron in the ISM of the outer disk by a factor of 2--3 between
1--3 Gyr. Once again, the solar radius is less sensitive to the IM-enhanced
IMF because the amplitude of the IM-bump decays on a shorter timescale.

As well as being SN~Ia progenitors, AGBs leave behind white dwarf
remnants that can be indirectly detected through microlensing
experiments. We found that the IM-enhanced IMF increases the
present-day number of WD remnants by only about 5\% and 10\% for the
solar and outer radii, respectively. Thus, the number of WDs is not
expected to be a sensitive discriminant of the different IMF
scenarios. This echos the Galactic halo results of Gibson \& Mould
(1997) and Fields, Freese, \& Graff (2000), who found that stellar C
and N abundances provide far stronger constraints than WD counts on
any enhancement in the number of LIMS at early times.

It should be noted that although \citetalias{AshenfelterT_04b} only
predict the evolution at the solar radius, if their IM-enhanced model
were applied to the outer disk we expect that they would \emph{not}
predict an increase in either the SN~Ia rate or [Fe/H] because their
SN~Ia prescription prohibits the formation of Type Ia SNe in metal-poor
environments.

\subsection{Magnesium isotopic ratio}\label{mg_results}

\begin{figure}
\centering
\includegraphics[width=8cm]{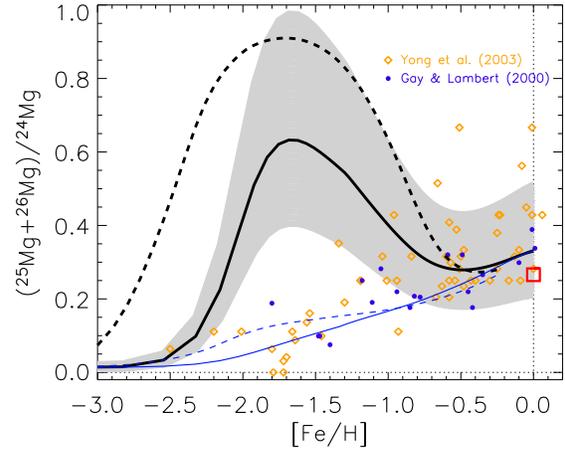}
      \caption{Evolution of ($^{25}$Mg+$^{26}$Mg)/$^{24}$Mg as a
      function of [Fe/H]. As for Fig~\ref{snia_fig}, the solid and
      dashed lines show predictions for the solar radius and outer
      disk, respectively. Thin lines denote a normal IMF model and
      thick lines represent the model with an IM-enhanced
      IMF. Circles and diamonds show stellar data from Gay \& Lambert
      (2000) and Yong et al. (2003b), respectively. The square shows
      the solar value. The shaded area indicates a $\pm$0.2 dex
      uncertainty range for the thick solid line.}  \label{mgiso_fig}
\end{figure}

Figure~\ref{mgiso_fig} shows the sensitivity of
($^{25}$Mg+$^{26}$Mg)/$^{24}$Mg versus [Fe/H] to the choice of IMF for
both the solar (8.5 kpc) and outer (16.5 kpc) radius
models. Predictions are plotted against abundances observed in nearby
dwarf stars by Gay \& Lambert (2000; circles) and Yong et al. (2003b;
diamonds). It has been suggested that the production of heavy Mg
isotopes by IMSs is needed to explain the observations in metal-poor
stars (e.g. Timmes et al. 1995; Goswami \& Prantzos 2000; Alib{\' e}s
et al. 2001). Indeed, Fenner et al. (2003) found that IMSs were
responsible for most of the heavy Mg isotopes in the solar
neighbourhood for ${\rm [Fe/H]}\!<\!-1$. The inclusion of AGB
nucleosynthesis within the framework of the standard IMF model
provides a good match to both datasets at low [Fe/H], as illustrated
by the thin solid line in Fig.~\ref{mgiso_fig}. At higher [Fe/H], the
normal IMF solar model matches the Gay \& Lambert (2000) data but not
those of Yong et al. (2003b). We caution that our predictions are best
compared against the Gay \& Lambert (2000) sample because Yong et
al. (2003b) used kinematics to preferentially select halo and thick
disk stars. Consequently, their sample contains relatively few thin
disk members and perhaps some stars belonging to an accreted component
(although inspection of figure 13 from Yong et al. 2003 indicates that
the fraction of accreted stars is small if one employs the Gratton et
al.  2003 criteria to specify the accreted component).  An uncertainty
in the $^{25}$Mg and $^{26}$Mg yields of $\pm$0.2 dex is illustrated
for the IM-enhanced solar radius model by the shaded region in
Fig.~\ref{mgiso_fig}, and will be discussed further in
Section~\ref{mg_uncertainties}.

Although the mean ($^{25}$Mg+$^{26}$Mg)/$^{24}$Mg for all stars with
${\rm [Fe/H]}\!<\!-1$ is about 0.15, Fig.~\ref{alpha_iso_fig}
demonstrates that a value 5--9 times larger than this is required to
explain the \da\ measured by MFW04, under the assumption of a solar
$^{25}$Mg:$^{26}$Mg ratio). \citetalias{AshenfelterT_04b} estimated
that a ratio of ($^{25}$Mg+$^{26}$Mg)/$^{24}$Mg = 0.62 would remove
the need for any time-variation in $\alpha$. This is lower than our
calculated value of 1.1 $\pm$ 0.3 (see Fig.~\ref{alpha_iso_fig})
because theirs is only a rough approximation and ours is based on the
QSO absorption-line spectra. However, to aid comparison, we followed
\citetalias{AshenfelterT_04b} in imposing an IM-enhanced IMF capable of
elevating ($^{25}$Mg+$^{26}$Mg)/$^{24}$Mg to $\sim$0.62 for the solar
radius model (thick solid line).

The behaviour of the Mg isotopic ratios in our IM-enhanced solar radius
model is very similar to \citetalias{AshenfelterT_04b} Model 1, reaching a
maximum of ($^{25}$Mg+$^{26}$Mg)/$^{24}$Mg = 0.63 at ${\rm
[Fe/H]}\!=\!-1.65$. The prolonged impact of the IM-enhanced IMF in the more
metal-poor outer disk leads to a maximum ($^{25}$Mg+$^{26}$Mg)/$^{24}$Mg
ratio of 0.92 at ${\rm [Fe/H]}\!=\!-1.7$. This is seven times higher than
the corresponding ratio in the case of the normal IMF. If we assume that
(i) the level of AGB stellar enhancement is a function of gas metallicity
and declines with the build-up of metals, and (ii) that many QSO
absorption systems consist of slowly evolving objects such as dwarfs or
outer disks of spirals, then these findings are qualitatively consistent
with a scenario in which much higher neutron-rich Mg isotopic abundances
are found in QSO absorption systems than in local stars of the same
metallicity.

\subsection{Nitrogen}\label{n_results}

\begin{figure}
\centering
\includegraphics[width=8cm]{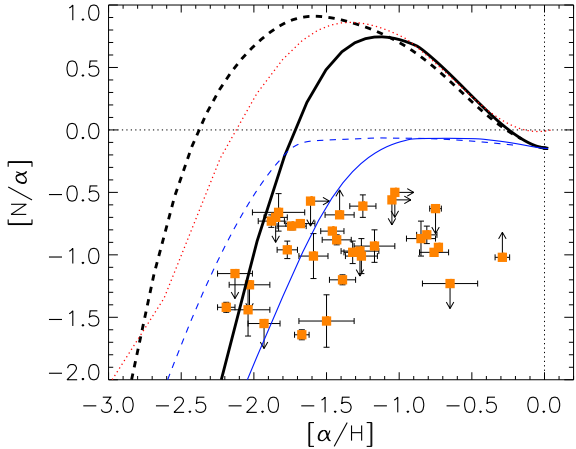}
      \caption{Evolution of [N/$\alpha$] as a function of
      [$\alpha$/H].  Dashed and solid lines have the same meaning as
      in Fig~\ref{snia_fig}. Model results employ Si as the reference
      $\alpha$-element. Data points with limits and error bars show
      DLA measurements by Centuri{\' o}n et al. (2003), where the
      $\alpha$-element is either O, Si, or S. The thin dotted line
      shows the IM-enhanced solar radius model in the case of van den
      Hoek \& Groenewegen (1997) IMS yields rather than Karakas \&
      Lattanzio (2003), to give an indication of the uncertainties
      associated with N production in the stellar models. } 
      \label{nonsi_fig}
\end{figure}

The types of AGB stars responsible for expelling enough heavy Mg isotopes
at early times to explain an apparent variation in $\alpha$ are also
thought to be the main factories of primary nitrogen. Thus, an enhanced AGB
population should leave a strong imprint on the N abundances in local stars
and in DLAs. The predicted variation of [N/$\alpha$] with [$\alpha$/H] is
plotted against DLA measurements in Fig.~\ref{nonsi_fig}.  The model
results are plotted using Si as the reference element, while for the
Centuri{\' o}n et al. (2003) data, the $\alpha$-element is O, Si or S. It
is reassuring that the normal IMF solar model (thin solid line) passes
through the DLA data before rising to roughly solar value. However, no
single homogeneous chemical evolution model will account for the broad
spread in the DLA data and, worryingly, the outer disk normal IMF model
(thin dashed line) overproduces N with respect to the data. Since the
normal IMF models (thin lines) already tend to produce more N than is
observed in DLAs, it is not surprising that the IM-enhanced IMF models
(thick lines) overproduce N by more than an order of magnitude. An
indication of the sensitivity of these results to the stellar
nucleosynthesis models is given by comparing the thick solid line with the
thin dotted line, which corresponds to the IM-enhanced solar radius model
in the case of van den Hoek \& Groenewegen (1997) IMS
yields. Section~\ref{cn_uncertainties} describes the uncertainties in these
predictions in greater detail.

\begin{figure}
\centering
\includegraphics[width=8cm]{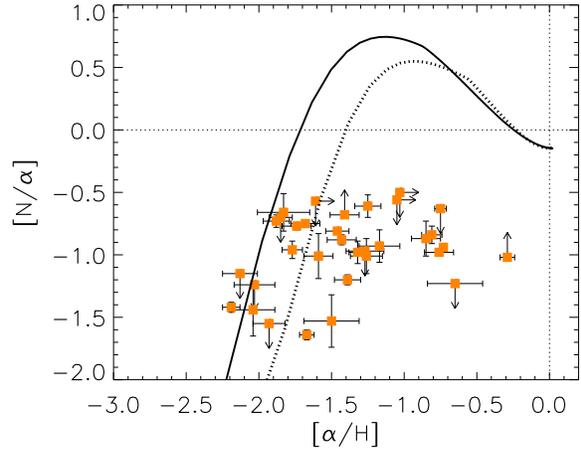}
      \caption{Sensitivity of [N/$\alpha$] vs [$\alpha$/H] to the star
      formation efficiency. The solid line represents the IM-enhanced
      solar radius model, as seen in Fig~\ref{nonsi_fig}. The dotted
      line shows the effect of increasing star formation (SF)
      efficiency by a factor of 2.5.  Increased SF efficiency improves
      the agreement with the DLA data (symbols), but still
      overproduces N at [$\alpha$/H] $\sim$ -1.}
      \label{nsfe_fig}
\end{figure}

Inspection of Figs.~\ref{mgiso_fig} \& \ref{nonsi_fig} reveals that the
empirical constraints imposed by DLA nitrogen abundances are strongly
violated by the models capable of producing sufficient $^{25,26}$Mg
relative to $^{24}$Mg to mimic the variation in the $\alpha$ obtained by
MFW04.  \citetalias{AshenfelterT_04b} encountered the same problem with
their Model 1, which they sought to rectify in Model 2 by increasing star
formation efficiency (SFE) by a factor of $\sim$2.5. Figure~\ref{nsfe_fig}
shows how our [N/$\alpha$] versus [$\alpha$/H] varies with a factor of 2.5
increase in SFE. The agreement with the data is improved, but [N/$\alpha$]
is still too high, except at the lowest metallicities.

\begin{figure}
\centering
\includegraphics[width=8cm]{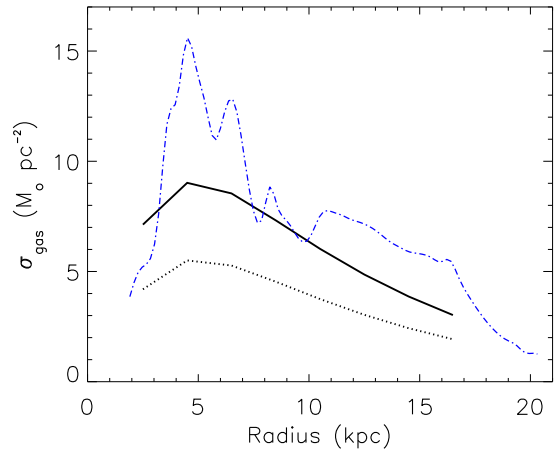}
      \caption{Present-day gas surface density profile. The dotted
      line presents the result for the model with star formation
      efficiency increased by a factor of 2.5 with respect to the
      standard model (solid line). The dot-dashed line corresponds to
      data from Dame (1993) based on observations of atomic and
      molecular hydrogen. The standard model replicates the
      observations to within a factor of two, whereas increased star
      formation efficiency leads to excess depletion of interstellar
      gas.}  \label{gasdens_fig}
\end{figure}

The curve corresponding to our increased SFE model in Fig.~\ref{nsfe_fig}
is extremely similar to the dot dash line in figure 15 from
\citetalias{AshenfelterT_04b}. We note that their Models 1 and 2 differ not
just in SFE, but in the location, width, amplitude, and decay timescale of
the IMF IM-bump. Nevertheless, we wish to point out a troublesome
consequence of such an increase in SFE that is robust to those
differences. Figure~\ref{gasdens_fig} shows the present-day gas surface
density profile expected in the case of normal or enhanced SFE. The
standard model agrees with the data from Dame (1993) to within a factor of
two. However, a factor of 2.5 increase in the SFE leads to severe gas
depletion, in conflict with the observations. While the chemical evolution
model that applies to QSO absorption systems need not satisfy the
empirical constraints from the Milky Way, we caution that increasing star
formation efficiency in order to alleviate overproduction of N might not be
appropriate given the gas-rich nature of many absorption systems.

\subsection{$^{13}$C/$^{12}$C abundance ratio}\label{ciso_results}

\begin{figure}
\centering
\includegraphics[width=8cm]{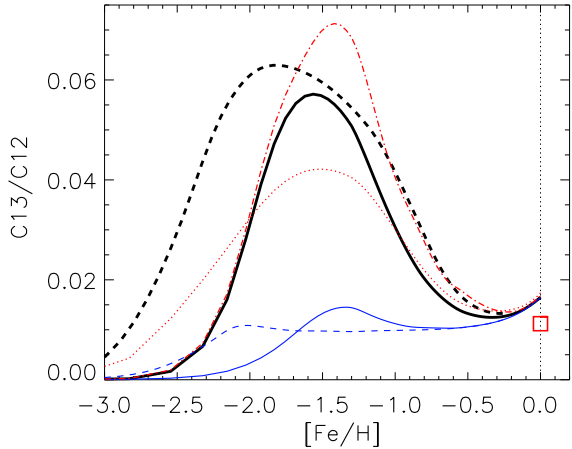}
      \caption{Evolution of $^{13}$C/$^{12}$C as a function of
      [Fe/H]. Symbols and lines have the same meaning as in
      Fig~\ref{snia_fig}.  The thin dotted and dot-dashed lines show
      the IM-enhanced solar radius model in the case of van den Hoek
      \& Groenewegen (1997) and Marigo (2001) IMS yields,
      respectively.  }
      \label{ciso_fig}
\end{figure}

Along with any significant contribution to the abundance of N and the heavy
Mg isotopes in metal-poor environments, low- and intermediate-mass stars
should leave an additional observable chemical signature in the form of
high $^{13}$C abundance relative to $^{12}$C. Metal-poor AGB stars with
mass $\sim$ 4 \msun\ are a major source of $^{13}$C, produced during hot
bottom burning when the CN cycle converts $^{12}$C into $^{13}$C. Since the
important factories of $^{12}$C are less massive 2--3\,\msun\ stars, the
$^{13}$C/$^{12}$C ratio in the ejecta of IMSs is strongly mass-dependent,
peaking sharply between 4 to 5 {\msun}.  Figure~\ref{ciso_fig} shows the
predicted evolution of $^{13}$C/$^{12}$C as a function of [Fe/H]. For both
the solar and outer radii models, the introduction of an IM-enhanced IMF
increases the $^{13}$C/$^{12}$C ratio 4--6-fold in the metallicity range
corresponding to the greatest enhancement of the heavy Mg isotopes. To
estimate the uncertainty in these results due to the stellar yields, we
also ran identical models using the LIMS yields from van den Hoek \&
Groenewegen (1997) and Marigo (2001), depicted with thin dotted and
dot-dashed lines, respectively. High values of $^{13}$C/$^{12}$C may
conflict with the symmetry of C{\sc \,iv} line profiles observed in some
QSO absorbers and could impose a further constraint on the shape of the IMF
at early times. This will be discussed in more detail in
Section~\ref{conclusions}.

\subsection{Silicon, aluminium and phosphorus}\label{sialp_results}

\begin{figure}
\centering
\includegraphics[width=8cm]{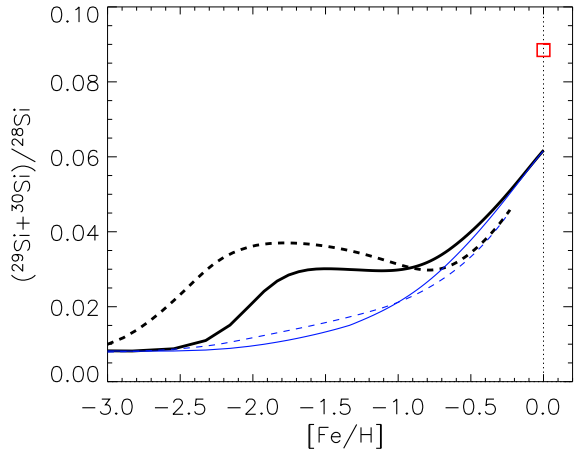}
      \caption{Evolution of ($^{29}$Si+$^{30}$Si)/$^{28}$Si as a
      function of [Fe/H]. Symbols and lines have the same meaning as in
      Fig~\ref{snia_fig}.}
      \label{siiso_fig}
\end{figure}

Since transitions from Si are important anchor lines in the many-multiplet
method, we show the predicted behaviour of ($^{29}$Si+$^{30}$Si)/$^{28}$Si
as a function [Fe/H] in Fig.~\ref{siiso_fig}. The relative abundance of Si
isotopes is not affected by an enhanced early population of AGB stars to
the extent of N and the heavy Mg and C isotopes. Indeed,
($^{29}$Si+$^{30}$Si)/$^{28}$Si remains well below the solar value (open
square) for both the local and outer disk models (thick
lines). Nevertheless, the \da\ measured in high-\zab\ QSO absorption
systems is largely independent of the Si isotopic composition, as
highlighted in Section~\ref{ss:AGB}. Thus, even a dramatic elevation in
($^{29}$Si+$^{30}$Si)/$^{28}$Si due to AGB stars would not account for the
high-redshift results from \citetalias{MurphyM_04a}.

\begin{figure}
\centering
\includegraphics[width=8cm]{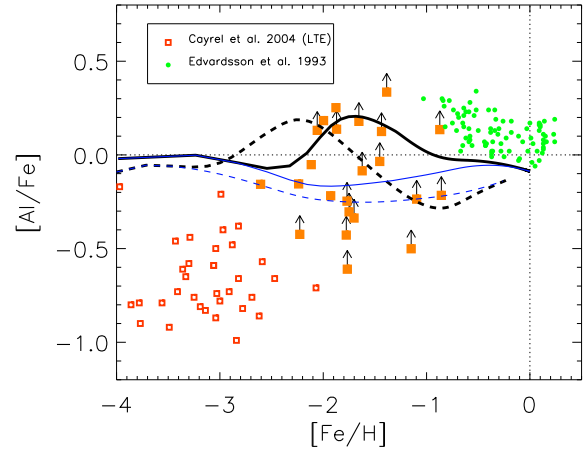}
      \caption{Variation of [Al/Fe] as a function of [Fe/H].  Dashed
      and solid lines have the same meaning as in
      Fig~\ref{snia_fig}. DLA measurements from Prochaska et al. are
      plotted with solid square symbols, while circles and open
      squares show stellar observations from Edvardsson et al. (1993)
      and Cayrel et al. (2004), respectively.
      }
      \label{al_fig}
\end{figure}

Aluminium is another element whose production by AGB stars is expected
to be important (e.g. Karakas \& Lattanzio 2003). The synthesis of Al
via the Mg-Al chain is particularly efficient in metal-poor
4--6\,\msun\ stars. The evolution of [Al/Fe] with [Fe/H] in the ISM is
depicted in Fig.~\ref{al_fig}. Observations of Al in both DLAs (solid
squares) and nearby stars (circles and open squares) covering a wide
range of metallicities make this element a potentially useful
discriminant of the different IMF models. Unfortunately, most of the
DLA measurements are lower limits, owing to line saturation (Prochaska
\& Wolfe 2002). The normal IMF models (thin lines) pass through the
lowermost DLA \emph{detections}, but only the IM-enhanced IMF models
(thick lines) can satisfy the highest DLA upper limits. The power of
Al to constrain the shape of the early IMF is diminished by the
inability of either solar radius model to match the stellar
observations at high and low metallicity. The overproduction of Al
with respect to metal-poor halo stars and the underproduction relative
to thin-disk stars is a problem of standard Galactic chemical
evolution models that has been noted by other authors (e.g. Timmes et
al. 1995; Alib{\' e}s et al. 2001). However, we note that local
thermodynamic equilibrium (LTE) calculations underestimate the Al
abundance in metal-poor stars. The open squares in Fig.~\ref{al_fig}
would be about 0.6 dex higher if non-LTE effects were included
(e.g. Gehren et al. 2004).

\begin{figure}
\centering
\includegraphics[width=8cm]{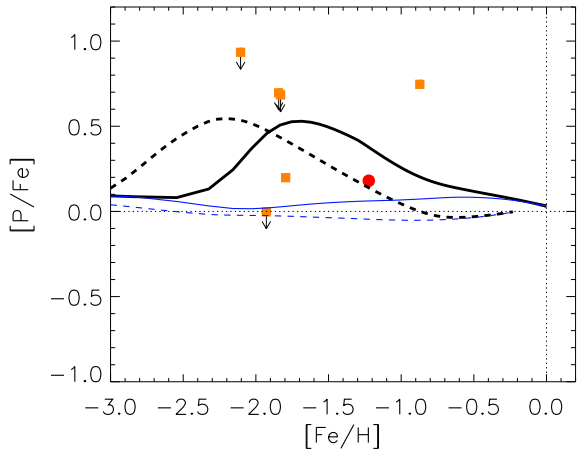}
      \caption{Variation of [P/Fe] as a function of [Fe/H].  Dashed
      and solid lines have the same meaning as in
      Fig~\ref{snia_fig}. DLA measurements from Prochaska et al. are
      plotted with square symbols, while the circle is from Outram,
      Chaffee, \& Carswell (1999).}  \label{p_fig}
\end{figure}

The production of phosphorus from IMSs leads to a $\sim$0.5~dex increase in
[P/Fe] when the IM-bump is added to the standard IMF, as shown in
Fig.~\ref{p_fig}. Stars in the 4--6\,\msun\ mass range are the chief
culprits for the additional $^{31}$P, which is understood to be generated
through neutron-capture onto Si followed by $\beta$-decay to $^{31}$P
(J. Lattanzio, private communication). As is the case for Al, the
difference between the two IMF models is significant, but the DLA
measurements do not allow the elimination of either scenario. In the case
of P, the DLA data is very scant and consists mostly of upper limits since
the P{\sc \,ii} line is often blended in the Ly$\alpha$ forest
(Dessauges-Zavadsky et al. 2004). Nevertheless, we present our predictions
for Al and P in the event that the growing collection of DLA measurements
may eventually provide a more stringent test of the models presented in
this paper.

\subsection{Uncertainties}\label{uncertainties}
\subsubsection{Magnesium}\label{mg_uncertainties}

Denissenkov \& Herwig (2003) and Denissenkov \& Weiss (2004) showed
that the production of $^{25}$Mg and $^{26}$Mg in a typical 5 \msun\
AGB star is fairly robust to the number of thermal-pulses, the HBB
temperature, and the third dredge-up efficiency. The final envelope
abundance of $^{25}$Mg and $^{26}$Mg in their series of models agreed
within $\sim$ 0.1 dex and $\sim$ 0.2 dex, respectively. Fenner et
al. (2004) estimated the sensitivity of the AGB yields of $^{25}$Mg
and $^{26}$Mg to the mass-loss prescription, finding them reduced by
$\sim$ 0.2 dex when the Vassiliadis \& Wood (1993) mass-loss law was
replaced with a steadier rate from Reimers (1975) (with $\eta = 3.5$
on the AGB). The treatment of convection in AGB models has recently
been investigated by Ventura (2004), who found that the yields were
very sensitive to the adoption of standard mixing length theory (MLT)
versus the full spectrum of turbulence (FST) model. The latter case
results in greater mass-loss rates, shorter AGB lifetimes, less third
dredge-up, and reduced yields of heavy elements including CNO and the
Mg isotopes. The FST model faces serious hurdles of its own, since it
predicts a deficit of Na in AGB ejecta and a \textit{positive} O-Na
correlation: at odds with observations of globular cluster stars
believed to be polluted by AGBs (Ventura, D'Antona, \& Mazzitelli
2004).

A further uncertainty relates to the presence of a $^{13}$C pocket,
which is neglected in the AGB models implemented in this study.  One
might expect that the inclusion of this additional source of neutrons
would increase the production of $^{25}$Mg and $^{26}$Mg. However the
impact of a $^{13}$C pocket on nucleosynthesis in metal-poor IMS
should be marginal in comparison to the effects of HBB (M. Lugaro,
\textit{private communication}).  Despite these unknowns, the ability
of our standard Galactic chemical evolution model to reproduce the Mg
isotopic ratios in metal-poor stars by adding in the Karakas \&
Lattanzio (2003) yields, suggests that their Mg predictions are
reasonably accurate.

The total production of $^{25,26}$Mg with respect to $^{24}$Mg from
Type~II SNe is uncertain to a similar degree due to internal
uncertainties in the stellar models as well as sensitivity to the
shape and upper mass limit of the IMF. Fenner et al. (2003) found that
a Salpeter (1955) IMF leads to 50\,per cent higher present-day
$^{26}$Mg/$^{24}$Mg ratio than in the case of a Kroupa et~al. (1993)
function, because the Salpeter law gives rise to a greater fraction of
massive stars (see their figs.~5 and 6). The shaded region in
Fig.~\ref{mgiso_fig} indicates the $\pm$ 0.2 dex uncertainty in the
results due to errors in both the AGB and SN~II contribution.

\subsubsection{Nitrogen and  $^{13}$C/$^{12}$C}\label{cn_uncertainties}

To estimate the uncertainty in the predicted evolution of N and the
$^{13}$C/$^{12}$C ratio, we ran identical models using two alternative
sets of IMS yields: those of van den Hoek \& Groenewegen (1997) and
Marigo (2001). The Marigo yields for N were in close agreement with
those of Karakas \& Lattanzio (2003) while the implementation of the
van den Hoek \& Groenewegen (1997) yields produced even more N. The
dotted line in Fig.~\ref{nonsi_fig} displays the results from an
IM-enhanced solar radius model using the van den Hoek \& Groenewegen
(1997) yields in place of Karakas et al. (i.e. compare dotted with
thick solid curve). It is clear that the problem of excess nitrogen is
further exacerbated in this case.

The dotted and dot-dashed lines in Fig.~\ref{ciso_fig} show the behaviour
of $^{13}$C/$^{12}$C assuming van den Hoek \& Groenewegen (1997) and Marigo
(2001) IMS yields, respectively. The Marigo yields generate the highest
$^{13}$C/$^{12}$C peak [with ($^{13}$C/$^{12}$C)$_{max}$ = 0.073] and those
of van den Hoek \& Groenewegen (1997) the lowest [with
($^{13}$C/$^{12}$C)$_{max}$ = 0.042]. The region encompassed by these two
extreme models is indicative of the level of uncertainty afflicting the
models.


\section{Discussion}\label{discussion}

\subsection{Summary of Results}

\citet{MurphyM_01b} pointed out that \da\ measurements are sensitive to the
isotopic composition of Mg in the gas-phase of the low-$z$ absorbers used
in varying-$\alpha$ studies.  Standard models of chemical evolution predict
that the abundance of the neutron-rich isotopes of Mg relative to $^{24}$Mg
decreases with lookback time, making the \citetalias{MurphyM_04a} \da\
result even more significant.  However, \citetalias{AshenfelterT_04b}
demonstrated that a sufficiently enhanced early population of
intermediate-mass stars can raise ($^{25}$Mg+$^{26}$Mg)/$^{24}$Mg to the
supersolar levels needed to render the \citetalias{MurphyM_04a} \da\ result
null.

A major problem with this scenario is the almost inevitable overproduction
of N when compared against the DLA data (Fig.~\ref{nonsi_fig}). This is
because the AGB stars responsible for significant $^{25,26}$Mg production
are also important sources of N. There are various ways to reconcile an
IM-enhanced IMF with the low N abundances in DLAs, but none are entirely
satisfactory. For instance, since nucleosynthesis within very metal-poor
IMSs is not well understood, the theoretical N yields may be
overestimated. However, in this case, it would be difficult to explain the
high N abundance found in some of the most iron-depleted stars in our
galaxy (e.g. Norris, Beers \& Ryan 2000; Christlieb et al. 2004).  Another
way to mitigate the problem of excess [N/$\alpha$] is through increased
star formation efficiency, however this leads to severe gas depletion that
may be inconsistent with the gas-rich nature of DLAs
(Fig.~\ref{gasdens_fig}).  Finally, the history of many DLAs may be marked
by periods of supernova-driven outflows, but it is not clear how such
galactic winds could preferentially remove N but not the heavy Mg isotopes.
Thus, the observed N abundance in QSO absorbers still stands as a robust
test of a putative IM-enhanced early IMF.

The enhanced $^{13}$C/$^{12}$C ratio predicted by these types of
chemical evolution models (Fig.~\ref{ciso_fig}) holds promise as a
future probe of the AGB contribution to QSO absorption line
abundances.  \citet{CarlssonJ_95a} has calculated the isotopic shifts
in the ubiquitous C{\sc \,iv} $\lambda$1548/1550 alkali doublet,
finding the $^{14}$C line to lie $\Delta v\!=\!10.3\,\kms$ to the blue
of the $^{12}$C line. Recent calculations by Berengut et al.~(in
preparation) confirm these large shifts to within 1\,per cent
precision and show the $^{13}$C--$^{12}$C separation to be $\Delta
v\!=\!5.5\,\kms$. Enhancements as large as $^{13}{\rm C}/^{12}{\rm
C}\!\approx\!0.1$ may already be ruled out by the symmetric C{\sc
\,iv} line profiles observed in $R\!\ga\!45\,000$ (${\rm
FWHM}\!\la\!6.7\,\kms$) spectra of highly ionized QSO absorbers with
simple velocity structure \citep[e.g.][]{PetitjeanP_04a}. However,
placing limits at lower $^{13}$C abundances may be unconvincing unless
the laboratory wavelengths of the C{\sc \,iv} doublet transitions can
be measured with better precision than 0.4\,\kms\
\citep{GriesmannU_00a}. If the calculated isotopic separations for
transitions of other C species (e.g.~C{\sc \,i} and C{\sc \,ii}) prove
to be as large as those for C{\sc \,iv}, they may provide a more
reliable probe of the $^{13}$C/$^{12}$C ratio in gas more closely
related to DLAs and, therefore, the chemical evolution models
presented here.

Given that many QSO absorbers are thought to probe slowly evolving
systems such as dwarf galaxies and the outer regions of galactic
disks, we simulated chemical evolution at large galactocentric radii,
where metal-enrichment is more gradual. We found that the chemical
signature from an enhanced IMS population at low metallicities was
more pronounced in the outer disk, with respect to the solar
neighbourhood. This follows directly from the assumed
metallicity-dependence of the IMS enhancement. Thus, our IM-enhanced
models predict that systems with the slowest build-up of metals will
bear the strongest signature of AGB-pollution.

The chemical abundance constraints imposed by QSO absorption systems do not
support excess numbers of AGB stars being formed in the types of systems
from which \citetalias{MurphyM_04a} derived a varying $\alpha$. Moreover, a
significantly enhanced AGB population in the local Galaxy is ruled out by
the sub-solar ratios of the neutron-rich Mg isotopes to $^{24}$Mg in nearby
metal-poor stars. We now summarise other arguments for and against the IMF
having a non-standard shape in the early universe.

\subsection{Other arguments for and against an early, IM-enhanced IMF}

\subsubsection{Observations of deuterium, carbon and nitrogen}

Motivated by the large observed scatter and possible trends with
metallicity of local and high-\zab\ D abundance determinations, Fields et
al. (2001) and Prantzos \& Ishimaru (2001) speculated that D could be
destroyed within stars without an accompanying increase in metals, provided
that: (i) the earliest populations of stars were strongly enhanced in IMSs,
and (ii) zero-metallicity IMSs do not release their synthesized C and N.
Prantzos \& Ishimaru (2001) emphasize that if the second provision is not
met, C and N abundances should be highly enhanced in high-\zab\ systems if
their D has been depleted through astration alone. They note that this is
inconsistent with the DLA data. Indeed, based on the N/$\alpha$ trend in
DLAs, Prochaska et al. (2002) presented a case for a top-heavy IMF, with
the birth of IMSs actually \emph{suppressed} in many QSO absorption systems.

The abundance of C and N in Galactic stars and DLAs provides one of the
strongest constraints on any enhancement in the population of LIMS in the
early Galaxy (Gibson \& Mould 1997; Fields, Freese, \& Graff
2000). However, this is contingent upon how much C and N is released by
extremely metal-poor IMSs. Fujimoto et al. (2000) provide evidence
favouring the suppression of C and N yields from zero-metallicity IMSs,
whereas Chieffi et al. (2001) reach the opposite conclusion. Abia et
al. (2001) propose that an early IMF peaked between 3--8 \msun\ helps
explain the presence of metal-poor Galactic stars with enhanced C and N
abundances, presuming that LIMS release C and N in large quantities. Thus,
the stellar models do not offer a cohesive picture of the level of C and N
enrichment from metal-poor LIMS, nor do observations of DLAs and metal-poor
stars clarify the situation. However, if AGBs are responsible for
$^{25,26}$Mg enhancements large enough to mimic a varying $\alpha$ in the
redshift range $0.5\!\la\!\zab\!\la\!1.8$, they will tend to have had a
much higher initial metallicity than primordial. Thus, their yields are not
subject to the very large uncertainties afflicting zero-metallicity models.

%
%
%

\vspace{-0.3cm}

\subsubsection{White dwarfs}

Evidence from microlensing surveys initially led to speculation that white
dwarf remnants contribute up to half of the mass of the Milky Way halo
(e.g. Bennett et al. 1996; Alcock et al. 1995). Populating the halo with
such a large fraction of white dwarfs requires the early IMF to be heavily
biased toward stars with initial mass in the range 1--8\,\msun.  Adams \&
Laughlin (1996) constructed IMFs capable of producing a white-dwarf
dominated halo, and \citetalias{AshenfelterT_04b} adopted IMF parameters
that were consistent with their deduced constraints. However, the white
dwarf contribution to the dynamical mass of the Galactic halo has since
been drastically revised downwards to less than a few percent (Gibson \&
Mould 1997; Flynn, Holopainen \& Holmberg 2003; Garcia-Berro et al. 2004;
Lee et al. 2004), obviating the need to invoke non-standard IMFs.

\vspace{-0.3cm}

\subsubsection{Theoretical star formation}

 Theoretical models of star formation suggest that the IMF of
 primordial gas would be biased toward higher mass stars (e.g. Abel,
 Bryan \& Norman 2000; Kroupa 2001; Hernandez \& Ferrara 2001; Mackey,
 Bromm \& Hernquist 2003; Clarke \& Bromm 2003). This is due to
 diminished cooling efficiency in metal-poor gas. However, recent
 hydrodynamical simulations of the collapse and fragmentation of
 primordial gas have predicted bimodal primordial IMFs with peaks at
 about 100 and 1 {\msun} (Nakamura \& Umemura 2001) and 30 and 0.3
 {\msun} (Omukai \& Yoshii 2003).

\vspace{-0.3cm}

\subsubsection{Extragalactic observations}

 Empirical support exists for an early IMF skewed toward high stellar
 masses in extragalactic environments. For instance, to explain the high
 metal content of the intra-cluster medium various top-heavy IMFs have been
 proposed (e.g. Gibson \& Matteucci 1997; Elbaz, Arnaud \& Vangioni-Flam
 1995). The photometric properties of ellipticals have also been
 well-explained assuming a top-heavy IMF (Arimoto \& Yoshii 1987).
 \citetalias{AshenfelterT_04b} note that their model is not inconsistent
 with the notion of a very early population of very massive stars that
 enrich the ISM to $Z\!\sim\!10^{-3}$.

\vspace{0.3cm}

In light of the uncertainties besetting theoretical simulations of
star formation and the interpretation of extragalactic photometric
data, careful analysis of the nucleosynthetic enrichment patterns of
QSO absorption systems may be a vital step in deciphering the shape of
the IMF in different astrophysical environments.


\section{Conclusions}\label{conclusions}

An early IMF heavily biased towards intermediate-mass stars is a potential
explanation of the $\alpha$-variation apparent in Keck/HIRES QSO
spectra. During their AGB phase these stars produce large quantities of
$^{25,26}$Mg which later pollute QSO absorption systems at
$\zab\!\la\!1.8$, causing an apparent shift in the Mg absorption lines and
a spurious $\da\!<\!0$. However, such a strong contribution from AGBs
should be evident in the detailed abundance patterns obtained in a growing
number of DLAs. Using chemical evolution models of a Milky Way-like spiral
galaxy at different radii, we have tested the idea that AGB stars were more
prevalent in the metal-poor Universe.  The predicted chemical ramifications
of an enhanced AGB population are not supported by the available data, with
the strongest argument against severe AGB pollution coming from the low
[N/$\alpha$] values observed in DLAs. Contrary to the conclusions of
\citetalias{AshenfelterT_04b}, we do not find consistent model parameters
which simultaneously explain the Keck/HIRES varying-$\alpha$ and abundance
data. We also contend that a variety of other arguments for invoking an
IM-enhanced IMF are, at best, unclear. Future measurements of (or limits
on) the $^{13}$C/$^{12}$C ratio from line-profile asymmetries and further
DLA observations of other elements synthesised by AGB stars, such as Al and
P, may provide additional constraints on the IMF's shape at early epochs.

The sensitivity of low-$z$ Mg/Fe \da\ measurements to isotopic abundance
variations in Mg (Fig.~\ref{alpha_iso_fig}) and the clear link to the
nucleosynthetic history of QSO absorbers demonstrates the importance of
careful \da\ measurements, even in the absence of real variations in
$\alpha$. If \da\ is measured to be zero in $\zab\!\la\!1.8$ absorbers via
independent means \citep[e.g.~by combining H{\sc \,i} 21-cm and molecular
absorption lines;][]{DarlingJ_03a} then future samples of Mg/Fe absorbers
might reveal Mg isotopic abundance variations in two ways: (i) through the
bulk relative line shifts between Mg and Fe transitions studied previously
and, (ii) if we assume that AGB pollution of the absorbers is not uniform,
through an increased scatter in the line shifts. Note that no increased
scatter is observed in the current HIRES or UVES samples
(\citetalias{MurphyM_04a}; \citealt{ChandH_04a}).


\section*{Acknowledgements}

The authors are extremely grateful to Maria Lugaro and Amanda Karakas
for many valuable discussions. We thank the anonymous referee for
comments that helped improve this paper. Financial support from the
Australian Research Council (ARC) is gratefully acknowledged. We also
acknowledge the Monash Cluster Computing Laboratory, the Victorian
Partnership for Advanced Computing and the Australian Partnership for
Advanced Computing for use of supercomputing facilities. MTM thanks
PPARC for support at the IoA under the observational rolling grant. YF
thanks the AFUW$-$SA for their support through the Daphne Elliot
Bursary.

\bspsmall

\label{lastpage}


\begin{thebibliography}{}
\small
\itemindent -0.48cm
\bibitem[Abel, Bryan, \& Norman(2000)]{2000ApJ...540...39A} Abel, T., 
Bryan, G.~L., \& Norman, M.~L.\ 2000, ApJ, 540, 39 

\bibitem[Abia et al.(2001)]{2001ApJ...557..126A} Abia, C., 
Dom{\'{\i}}nguez, I., Straniero, O., Limongi, M., Chieffi, A., \& Isern, 
J.\ 2001, ApJ, 557, 126 

\bibitem[Adams \& Laughlin(1996)]{1996ApJ...468..586A} Adams, F.~C.~\& 
Laughlin, G.\ 1996, ApJ, 468, 586 

\bibitem[Alcock et al.(1995)]{1995PhRvL..74.2867A} Alcock, C., et al.\ 
1995, Phys.~Rev.~Lett., 74, 2867 

\bibitem[Alib{\' e}s, Labay, \& Canal(2001)]{2001A&A...370.1103A} Alib{\'
e}s, A., Labay, J., \& Canal, R.\ 2001, A\&A, 370, 1103

\bibitem[Arimoto \& Yoshii(1987)]{1987A&A...173...23A} Arimoto, N.~\& 
Yoshii, Y.\ 1987, A\&A, 173, 23 

\bibitem[\protect\citeauthoryear{{Ashenfelter}, {Mathews} \&
  {Olive}}{{Ashenfelter} et~al.}{2004a}]{AshenfelterT_04a}
{Ashenfelter} T.~P., {Mathews} G.~J.,    {Olive} K.~A.,  2004a, Phys.~Rev.~Lett., 92, 041102

\bibitem[\protect\citeauthoryear{{Ashenfelter}, {Mathews} \&
  {Olive}}{{Ashenfelter} et~al.}{2004b}]{AshenfelterT_04b}
{Ashenfelter} T.~P.,  {Mathews} G.~J.,    {Olive} K.~A.,  2004b, ApJ, 615, 82

\bibitem[\protect\citeauthoryear{{Bethe} \& {Salpeter}}{{Bethe} \&
  {Salpeter}}{1977}]{BetheH_77a}
{Bethe} H.~A.,  {Salpeter} E.~E.,  1977, {Quantum mechanics of one- and
  two-electron atoms}.
Plenum, New York, NY, USA

\bibitem[Bennett et al.(1996)]{1996clfu.conf...95B} Bennett, D.~P., et al.\ 
1996, ASP Conf.~Ser.~ 88: Clusters, Lensing, and the Future of the 
Universe, 95 

\bibitem[\protect\citeauthoryear{{Berengut},
{Dzuba} \& {Flambaum}}{{Berengut} et~al.}{2003}]{BerengutJ_03a}{Berengut} J.~C.,
{Dzuba} V.~A.,    {Flambaum} V.~V.,  2003, Phys.~Rev.~A, 68, 022502

\bibitem[\protect\citeauthoryear{{Carlsson},
{Jonsson}, {Godefroid} \&  {Fischer}}{{Carlsson} et~al.}{1995}]{CarlssonJ_95a}
{Carlsson} J.,  {Jonsson} P.,  {Godefroid} M.~R.,    {Fischer} C.~F.,  1995,
  J.~Phys.~B, 28, 3729

\bibitem[Cayrel et al.(2004)]{2004A&A...416.1117C} Cayrel, R., et al.\ 
2004, A\&A, 416, 1117 

\bibitem[Centuri{\' o}n et al.(2003)]{2003A&A...403...55C} Centuri{\' o}n, 
M., Molaro, P., Vladilo, G., P{\' e}roux, C., Levshakov, S.~A., \& 
D'Odorico, V.\ 2003, A\&A, 403, 55

\bibitem[\protect\citeauthoryear{{Chand}, {Petitjean}, {Srianand} \&
  {Aracil}}{{Chand} et~al.}{2004b}]{ChandH_04b}
{Chand} H.,  {Petitjean} P.,  {Srianand} R.,    {Aracil} B.,  2004b, A\&A, 430, 47

\bibitem[\protect\citeauthoryear{{Chand}, {Srianand}, {Petitjean} \&
  {Aracil}}{{Chand} et~al.}{2004a}]{ChandH_04a}
{Chand} H.,  {Srianand} R.,  {Petitjean} P.,    {Aracil} B.,  2004a, A\&A, 417,
  853

\bibitem[Chiappini, Matteucci \& Gratton(1997)]{ChiappiniC_97a} Chiappini,
C., Matteucci, F. \& Gratton, R. 1997, ApJ, 477, 765

\bibitem[Chieffi, Dom{\'{\i}}nguez, Limongi, \& 
Straniero(2001)]{2001ApJ...554.1159C} Chieffi, A., Dom{\'{\i}}nguez, I., 
Limongi, M., \& Straniero, O.\ 2001, ApJ, 554, 1159 

\bibitem[Christlieb et al.(2004)]{2004ApJ...603..708C} Christlieb, N., 
Gustafsson, B., Korn, A.~J., Barklem, P.~S., Beers, T.~C., Bessell, M.~S., 
Karlsson, T., \& Mizuno-Wiedner, M.\ 2004, ApJ, 603, 708 

\bibitem[Clarke \& Bromm(2003)]{2003MNRAS.343.1224C} Clarke, C.~J.~\& 
Bromm, V.\ 2003, MNRAS, 343, 1224 

\bibitem[\protect\citeauthoryear{{Cowie} \& {Songaila}}{{Cowie} \&
  {Songaila}}{1995}]{CowieL_95a}
{Cowie} L.~L.,  {Songaila} A.,  1995, ApJ, 453, 596

\bibitem[Dame(1993)]{1993AIPC..278..267D} Dame, T.~M.\ 1993, AIP 
Conf.~Proc.~278: Back to the Galaxy, 278, 267 

\bibitem[\protect\citeauthoryear{{Darling}}{{Darling}}{2003}]{DarlingJ_03a}
{Darling} J.,  2003, Phys.~Rev.~Lett., 91, 011301

\bibitem[Denissenkov \& Herwig(2003)]{2003ApJ...590L..99D} Denissenkov,
    P.~A.~\& Herwig, F.\ 2003, ApJL, 590, L99

\bibitem[Denissenkov \& Weiss(2004)]{2004ApJ...603..119D} Denissenkov,
    P.~A.~\& Weiss, A.\ 2004, ApJ, 603, 119

\bibitem[Dessauges-Zavadsky et al.(2004)]{2004A&A...416...79D} 
Dessauges-Zavadsky, M., Calura, F., Prochaska, J.~X., D'Odorico, S., \& 
Matteucci, F.\ 2004, A\&A, 416, 79 

\bibitem[\protect\citeauthoryear{{Dirac}}{{Dirac}}{1937}]{DiracP_37a}
{Dirac} P.~A.~M.,  1937, Nat, 139, 323

\bibitem[\protect\citeauthoryear{{Drullinger}, {Wineland} \&
  {Bergquist}}{{Drullinger} et~al.}{1980}]{DrullingerR_80a}
{Drullinger} R.~E.,  {Wineland} D.~J.,    {Bergquist} J.~C.,  1980,
Appl.~Phys., 22,  365

\bibitem[\protect\citeauthoryear{{Dzuba}, {Flambaum} \& {Webb}}{{Dzuba}
  et~al.}{1999a}]{DzubaV_99b}
{Dzuba} V.~A.,  {Flambaum} V.~V.,    {Webb} J.~K.,  1999a, Phys.~Rev.~A, 59, 230

\bibitem[\protect\citeauthoryear{{Dzuba}, {Flambaum} \& {Webb}}{{Dzuba}
  et~al.}{1999b}]{DzubaV_99a}
{Dzuba} V.~A.,  {Flambaum} V.~V.,    {Webb} J.~K.,  1999b, Phys.~Rev.~Lett., 82, 888

\bibitem[Edvardsson et al.(1993)]{}Edvardsson, B., Andersen, J.,
Gustafsson, B., Lambert, D.~L., Nissen, P.~E., \& Tomkin, J.\ 1993,
A\&A, 275, 101

\bibitem[Elbaz, Arnaud, \& Vangioni-Flam(1995)]{1995A&A...303..345E} Elbaz, 
D., Arnaud, M., \& Vangioni-Flam, E.\ 1995, A\&A, 303, 345 

\bibitem[Fenner \& Gibson(2003)]{2003PASA...20..189F} Fenner, Y.~\& Gibson, 
B.~K.\ 2003, Publ.~Astron.~Soc.~Aust., 20, 189 

\bibitem[Fenner et al.(2003)]{2003PASA...20..340F} Fenner, Y., Gibson, 
B.~K., Lee, H.-c., Karakas, A.~I., Lattanzio, J.~C., Chieffi, A., Limongi, 
M., \& Yong, D.\ 2003, Publ.~Astron.~Soc.~Aust., 20, 340 

\bibitem[Fenner, Prochaska, \& Gibson(2004)]{2004ApJ...606..116F} Fenner, 
Y., Prochaska, J.~X., \& Gibson, B.~K.\ 2004, ApJ, 606, 116 

\bibitem[Fenner et al.(2004)]{2004MNRAS.tmp..280F} Fenner, Y., Campbell, 
S., Karakas, A.~I., Lattanzio, J.~C., \& Gibson, B.~K.\ 2004, MNRAS, 280 

\bibitem[Fields, Freese, \& Graff(2000)]{2000ApJ...534..265F} Fields, 
B.~D., Freese, K., \& Graff, D.~S.\ 2000, ApJ, 534, 265 

\bibitem[Fields, Mathews, \& Schramm(1997)]{1997ApJ...483..625F} Fields, 
B.~D., Mathews, G.~J., \& Schramm, D.~N.\ 1997, ApJ, 483, 625 

\bibitem[Fields et al.(2001)]{2001ApJ...563..653F} Fields, B.~D., Olive, 
K.~A., Silk, J., Cass{\' e}, M., \& Vangioni-Flam, E.\ 2001, ApJ, 563, 653 

\bibitem[Flynn, Holopainen, \& Holmberg(2003)]{2003MNRAS.339..817F} Flynn, 
C., Holopainen, J., \& Holmberg, J.\ 2003, MNRAS, 339, 817 

\bibitem[Fujimoto, Ikeda, \& Iben(2000)]{2000ApJ...529L..25F} Fujimoto, 
M.~Y., Ikeda, Y., \& Iben, I.~J.\ 2000, ApJL, 529, L25 

\bibitem[Garc{\'{\i}}a-Berro, Torres, Isern, \& 
Burkert(2004)]{2004A&A...418...53G} Garc{\'{\i}}a-Berro, E., Torres, S., 
Isern, J., \& Burkert, A.\ 2004, A\&A, 418, 53 

\bibitem[\protect\citeauthoryear{{Gay} \& {Lambert}}{{Gay} \&
  {Lambert}}{2000}]{GayP_00a}
{Gay} P.~L.,  {Lambert} D.~L.,  2000, ApJ, 533, 260

\bibitem[Gehren et al.(2004)]{2004A&A...413.1045G} Gehren, T., Liang, 
Y.~C., Shi, J.~R., Zhang, H.~W., \& Zhao, G.\ 2004, A\&A, 413, 1045 

\bibitem[Gibson et al.(2001)]{GibsonB_01a} Gibson, B.K., Giroux, M.L.,
Penton, S.V., Stocke, J.T., Shull, J.M. \& Tumlinson, J. 2001, AJ, 547,
3280

\bibitem[Gibson \& Matteucci(1997)]{1997MNRAS.291L...8G} Gibson, B.~K.~\& 
Matteucci, F.\ 1997, MNRAS, 291, L8 

\bibitem[Gibson \& Mould(1997)]{1997ApJ...482...98G} Gibson, B.~K.~\& 
Mould, J.~R.\ 1997, ApJ, 482, 98 

\bibitem[Goswami \& Prantzos(2000)]{GoswamiA_00a} Goswami, A. \& Prantzos,
N., 2000, A\&A, 359, 191

\bibitem[Gratton et al.(2003)]{Gratton_03} Gratton, R. G., Carretta, E., 
Claudi, R., Lucatello, S. \& Barbieri, M., 2003, A\&A, 404, 187

\bibitem[Greggio \& Renzini(1983)]{GreggioL_83a} Greggio, L.~\& Renzini,
A.\ 1983, A\&A, 118, 217

\bibitem[\protect\citeauthoryear{{Griesmann} \& {Kling}}{{Griesmann} \&
  {Kling}}{2000}]{GriesmannU_00a}
{Griesmann} U.,  {Kling} R.,  2000, ApJ, 536, L113

\bibitem[\protect\citeauthoryear{{Hallstadius}}{{Hallstadius}}{1979}]
  {HallstadiusL_79a}
{Hallstadius} L.,  1979, Z.~Phys.~A, 291, 203

\bibitem[Hernandez \& Ferrara(2001)]{2001MNRAS.324..484H} Hernandez, X.~\& 
Ferrara, A.\ 2001, MNRAS, 324, 484 

\bibitem[Iwamoto et al.(1999)]{1999ApJS..125..439I} Iwamoto, K., Brachwitz,
F., Nomoto, K., Kishimoto, N., Umeda, H., Hix, W.~R., \& Thielemann, F.\
1999, ApJS, 125, 439

\bibitem[Karakas \& Lattanzio(2003)]{2003PASA...20..279K} Karakas, A.~I.~\&
  Lattanzio, J.~C.\ 2003, Publ.~Astron.~Soc.~Aust., 20, 279

\bibitem[Kobayashi, Tsujimoto, \& Nomoto(2000)]{2000ApJ...539...26K} 
Kobayashi, C., Tsujimoto, T., \& Nomoto, K.\ 2000, ApJ, 539, 26 

\bibitem[\protect\citeauthoryear{{Kozlov}, {Korol}, {Berengut}, {Dzuba} \&
  {Flambaum}}{{Kozlov} et~al.}{2004}]{KozlovM_04a}
{Kozlov} M.~G.,  {Korol} V.~A.,  {Berengut} J.~C.,  {Dzuba} V.~A.,
  {Flambaum} V.~V.,  2004, Phys.~Rev.~A, 70, 062108

\bibitem[Kroupa, Tout, \& Gilmore(1993)]{1993MNRAS.262..545K} Kroupa, P.,
Tout, C.~A., \& Gilmore, G.\ 1993, MNRAS, 262, 545

\bibitem[Kroupa(2001)]{KroupaP_01a} Kroupa, P., 2001, MNRAS, 322, 231

\bibitem[Lee et al.(2004)]{2004PASA...21..153L} Lee, H., Gibson, B.~K., 
Fenner, Y., Brook, C.~B., Kawata, D., Renda, A., Holopainen, J., \& Flynn, 
C.\ 2004, Publ.~Astron.~Soc.~Aust., 21, 153 

\bibitem[\protect\citeauthoryear{{Levshakov}, {Centurion}, {Molaro} \&
  {D'Odorico}}{{Levshakov} et~al.}{2004}]{LevshakovS_04a}
{Levshakov} S.~A.,  {Centurion} M.,  {Molaro} P.,    {D'Odorico} S.,  2004,
  A\&A, submitted, preprint (astro-ph/0408188)

\bibitem[Mackey, Bromm, \& Hernquist(2003)]{2003ApJ...586....1M} Mackey, 
J., Bromm, V., \& Hernquist, L.\ 2003, ApJ, 586, 1 

\bibitem[Marigo(2001)]{2001A&A...370..194M} Marigo, P.\ 2001, A\&A, 370, 
194 

\bibitem[Matteucci \& Greggio(1986)]{MatteucciF_86a} Matteucci, F.~\&
Greggio, L.\ 1986, A\&A, 154, 279

\bibitem[\protect\citeauthoryear{{Milne}}{{Milne}}{1935}]{MilneE_35a}
{Milne} E.~A.,  1935, Relativity, Gravitation and World Structure.
Clarendon Press, Oxford, UK

\bibitem[\protect\citeauthoryear{{Milne}}{{Milne}}{1937}]{MilneE_37a}
{Milne} E.~A.,  1937, Proc.~R.~Soc.~A, 158, 324

\bibitem[\protect\citeauthoryear{{Murphy}, {Flambaum}, {Webb}, {Dzuba},
  {Prochaska} \& {Wolfe}}{{Murphy} et~al.}{2004}]{MurphyM_04a}
{Murphy} M.~T.,  {Flambaum} V.~V.,  {Webb} J.~K.,  {Dzuba} V.~V.,  {Prochaska}
  J.~X.,    {Wolfe} A.~M.,  2004, Lecture~Notes~Phys., 648, 131

\bibitem[\protect\citeauthoryear{{Murphy}, {Webb} \& {Flambaum}}{{Murphy}
  et~al.}{2003}]{MurphyM_03a}
{Murphy} M.~T.,  {Webb} J.~K.,    {Flambaum} V.~V.,  2003, MNRAS, 345, 609

\bibitem[\protect\citeauthoryear{{Murphy}, {Webb}, {Flambaum}, {Churchill} \&
  {Prochaska}}{{Murphy} et~al.}{2001b}]{MurphyM_01b}
{Murphy} M.~T.,  {Webb} J.~K.,  {Flambaum} V.~V.,  {Churchill} C.~W.,
  {Prochaska} J.~X.,  2001b, MNRAS, 327, 1223

\bibitem[\protect\citeauthoryear{{Murphy}, {Webb}, {Flambaum}, {Dzuba},
  {Churchill}, {Prochaska}, {Barrow} \& {Wolfe}}{{Murphy}
  et~al.}{2001a}]{MurphyM_01a}
{Murphy} M.~T.,  {Webb} J.~K.,  {Flambaum} V.~V.,  {Dzuba} V.~A.,  {Churchill}
  C.~W.,  {Prochaska} J.~X.,  {Barrow} J.~D.,    {Wolfe} A.~M.,  2001a, MNRAS,
  327, 1208

\bibitem[\protect\citeauthoryear{{Murphy}, {Webb}, {Flambaum}, {Prochaska} \&
  {Wolfe}}{{Murphy} et~al.}{2001c}]{MurphyM_01c}
{Murphy} M.~T.,  {Webb} J.~K.,  {Flambaum} V.~V.,  {Prochaska} J.~X.,
  {Wolfe} A.~M.,  2001c, MNRAS, 327, 1237

\bibitem[Nakamura \& Umemura(2001)]{2001ApJ...548...19N} Nakamura, F.~\& 
Umemura, M.\ 2001, ApJ, 548, 19 

\bibitem[Norris, Beers, \& Ryan(2000)]{2000ApJ...540..456N} Norris, J.~E., 
Beers, T.~C., \& Ryan, S.~G.\ 2000, ApJ, 540, 456

\bibitem[Omukai \& Yoshii(2003)]{2003ApJ...599..746O} Omukai, K.~\& Yoshii, 
Y.\ 2003, ApJ, 599, 746 

\bibitem[Outram, Chaffee, \& Carswell(1999)]{1999MNRAS.310..289O} Outram, 
P.~J., Chaffee, F.~H., \& Carswell, R.~F.\ 1999, MNRAS, 310, 289 

\bibitem[Pagel(2001)]{2001PASP..113..137P} Pagel, B.~E.~J.\ 2001, PASP, 
113, 137 

\bibitem[\protect\citeauthoryear{{Petitjean} \& {Aracil}}{{Petitjean} \&
  {Aracil}}{2004}]{PetitjeanP_04a}
{Petitjean} P.,  {Aracil} B.,  2004b, A\&A, 422, 523

\bibitem[Pettini, Ellison, Steidel, \& Bowen(1999)]{1999ApJ...510..576P} 
Pettini, M., Ellison, S.~L., Steidel, C.~C., \& Bowen, D.~V.\ 1999, ApJ, 
510, 576 

\bibitem[\protect\citeauthoryear{{Pickering}, {Thorne} \&
{Webb}}{{Pickering} et~al.}{1998}]{PickeringJ_98a}
{Pickering} J.~C.,  {Thorne} A.~P.,    {Webb} J.~K.,  1998, MNRAS, 300, 131

\bibitem[Prantzos \& Ishimaru(2001)]{2001A&A...376..751P} Prantzos, N.~\& 
Ishimaru, Y.\ 2001, A\&A, 376, 751 

\bibitem[Prochaska \& Wolfe(2002)]{2002ApJ...566...68P} Prochaska, J.~X.~\& 
Wolfe, A.~M.\ 2002, ApJ, 566, 68 

\bibitem[Prochaska et al.(2002)]{2002PASP..114..933P} Prochaska, J.~X., 
Henry, R.~B.~C., O'Meara, J.~M., Tytler, D., Wolfe, A.~M., Kirkman, D., 
Lubin, D., \& Suzuki, N.\ 2002, PASP, 114, 933 

\bibitem[\protect\citeauthoryear{{Quast}, {Reimers} \& {Levshakov}}{{Quast}
  et~al.}{2004}]{QuastR_04a}
{Quast} R.,  {Reimers} D.,    {Levshakov} S.~A.,  2004b, A\&A, 415, L7

\bibitem[Reimers(1975)]{1975MSRSL...8..369R} Reimers, D. 1975, Memoires of
  the Societe Royale des Sciences de Liege, 8, 369

\bibitem[Renzini \& Voli(1981)]{RenziniA_81a} Renzini, A. \& Voli, M. 1981,
A\&A, 94, 175

\bibitem[Romano et al.(2000)]{RomanoD_00a} Romano, D., Matteucci, F.,
  Salucci, P. \& Chiappini, C. 2000, ApJ, 539, 235

\bibitem[\protect\citeauthoryear{{Rosman} \& {Taylor}}{{Rosman} \&
  {Taylor}}{1998}]{RosmanK_98a}
{Rosman} K.~J.~R.,  {Taylor} P.~D.~P.,  1998, J.~Phys.~Chem.~Ref.~Data, 27, 1275

\bibitem[Salpeter(1955)]{1955ApJ...121..161S} Salpeter, E.~E.\ 1955, ApJ,
121, 161

\bibitem[Schaller, Schaerer, Meynet, \& Maeder(1992)]{1992A&AS...96..269S}
Schaller, G., Schaerer, D., Meynet, G., \& Maeder, A.\ 1992, A\&AS, 96, 269

\bibitem[Schmidt(1959)]{SchmidtM_59a} Schmidt, M. 1959, ApJ, 129, 243

\bibitem[Sembach et al.(2002)]{SembackK_02a} Sembach, K.R., Gibson, B.K.,
Fenner, Y. \& Putman, M.E.  2002, ApJ, 572, 178

\bibitem[Shetrone, C{\^ o}t{\' e}, \& Sargent(2001)]{2001ApJ...548..592S} 
Shetrone, M.~D., C{\^ o}t{\' e}, P., \& Sargent, W.~L.~W.\ 2001, ApJ, 548, 
592 

\bibitem[\protect\citeauthoryear{{Srianand}, {Chand}, {Petitjean} \&
  {Aracil}}{{Srianand} et~al.}{2004}]{SrianandR_04a}
{Srianand} R.,  {Chand} H.,  {Petitjean} P.,    {Aracil} B.,  2004, Phys.~Rev.~Lett., 92,
  121302

\bibitem[Thielemann, Nomoto, \& Yokoi(1986)]{1986A&A...158...17T}
Thielemann, F.-K., Nomoto, K., \& Yokoi, K.\ 1986, A\&A, 158, 17

\bibitem[\protect\citeauthoryear{{Timmes}, {Woosley} \& {Weaver}}{{Timmes}
  et~al.}{1995}]{TimmesF_95a}
{Timmes} F.~X.,  {Woosley} S.~E.,    {Weaver} T.~A.,  1995, ApJS, 98, 617

\bibitem[Tinsley(1980)]{TinsleyB_80a} Tinsley, B.M. 1980,
Fund. Cosm. Phys., 5, 287

\bibitem[\protect\citeauthoryear{{Uzan}}{{Uzan}}{2003}]{UzanJ_03a}
{Uzan} J.,  2003, Rev.~Mod.~Phys., 75, 403

\bibitem[van den Hoek \& Groenewegen(1997)]{1997A&AS..123..305V} van den
Hoek, L.~B.~\& Groenewegen, M.~A.~T.\ 1997, A\&AS, 123, 305

\bibitem[\protect\citeauthoryear{{Varshalovich}, {Panchuk} \&
  {Ivanchik}}{{Varshalovich} et~al.}{1996}]{VarshalovichD_96b}
{Varshalovich} D.~A.,  {Panchuk} V.~E.,    {Ivanchik} A.~V.,  1996, Astron.~Lett., 22, 6

\bibitem[\protect\citeauthoryear{{Varshalovich} \& {Potekhin}}{{Varshalovich}
  \& {Potekhin}}{1994}]{VarshalovichD_94a}
{Varshalovich} D.~A.,  {Potekhin} A.~Y.,  1994, Astron.~Lett., 20, 771

\bibitem[\protect\citeauthoryear{{Varshalovich}, {Potekhin} \&
  {Ivanchik}}{{Varshalovich} et~al.}{2000}]{VarshalovichD_00a}
{Varshalovich} D.~A.,  {Potekhin} A.~Y.,    {Ivanchik} A.~V.,  2000, in
  {Dunford} R.~W.,  {Gemmel} D.~S.,  {Kanter} E.~P.,  {Kraessig} B.,
  {Southworth} S.~H.,   {Young} L.,  eds, AIP Conf. Proc. Vol. 506, X-Ray and
  Inner-Shell Processes. Argonne National Laboratory, Argonne, IL, USA, p.~503

\bibitem[Vassiliadis \& Wood(1993)]{1993ApJ...413..641V} Vassiliadis, E.~\&
  Wood, P.~R.\ 1993, ApJ, 413, 641

\bibitem[Ventura(2004)]{2004MmSAI..75..654V} Ventura, P.\ 2004, Memorie 
della Societa Astronomica Italiana, 75, 654 

\bibitem[Ventura, D'Antona, \& Mazzitelli(2004)]{2004MmSAI..75..335V} 
Ventura, P., D'Antona, F., \& Mazzitelli, I.\ 2004, Memorie della Societa 
Astronomica Italiana, 75, 335 

\bibitem[Wakker et al.(1999)]{WakkerB_99a} Wakker B.P.~{et al.}, 1999,
Nat, 402, 388

\bibitem[\protect\citeauthoryear{{Webb}, {Flambaum}, {Churchill}, {Drinkwater}
  \& {Barrow}}{{Webb} et~al.}{1999}]{WebbJ_99a}
{Webb} J.~K.,  {Flambaum} V.~V.,  {Churchill} C.~W.,  {Drinkwater} M.~J.,
  {Barrow} J.~D.,  1999, Phys.~Rev.~Lett., 82, 884

\bibitem[\protect\citeauthoryear{{Webb}, {Murphy}, {Flambaum}, {Dzuba},
  {Barrow}, {Churchill}, {Prochaska} \& {Wolfe}}{{Webb}
  et~al.}{2001}]{WebbJ_01a}
{Webb} J.~K.,  {Murphy} M.~T.,  {Flambaum} V.~V.,  {Dzuba} V.~A.,  {Barrow}
  J.~D.,  {Churchill} C.~W.,  {Prochaska} J.~X.,    {Wolfe} A.~M.,  2001, Phys.~Rev.~Lett.,
  87, 091301

\bibitem[Woosley \& Weaver(1995)]{1995ApJS..101..181W} Woosley, S.~E.~\&
Weaver, T.~A.\ 1995, ApJS, 101, 181

\bibitem[\protect\citeauthoryear{{Yong}, {Grundahl}, {Lambert}, {Nissen} \&
  {Shetrone}}{{Yong} et~al.}{2003a}]{YongD_03a}
{Yong} D.,  {Grundahl} F.,  {Lambert} D.~L.,  {Nissen} P.~E.,    {Shetrone}
  M.~D.,  2003a, A\&A, 402, 985

\bibitem[\protect\citeauthoryear{{Yong}, {Lambert} \& {Ivans}}{{Yong}
  et~al.}{2003b}]{YongD_03b}
{Yong} D.,  {Lambert} D.~L.,    {Ivans} I.~I.,  2003b, ApJ, 599, 1357

\end{thebibliography}
\end{document}